\documentclass[11pt]{article}
\usepackage{amsfonts}
 \usepackage{pifont}
\usepackage{amsmath,amsfonts,amssymb,theorem,graphicx,listings,txfonts}
\usepackage{graphics}
 \usepackage{caption2}
\usepackage{flafter}
 \usepackage{indentfirst}
\usepackage{epsfig}
  \usepackage{mathrsfs}
\usepackage{subfigure}
 \usepackage{cite}
\newtheorem{theorem}{Theorem}[section]

\newtheorem{proposition}[theorem]{Proposition}
\newtheorem{corollary}[theorem]{Corollary}
\newtheorem{definition}[theorem]{Definition}
\newtheorem{algorithm}[theorem]{Algorithm}

\theoremstyle{example}
\newtheorem{example}[theorem]{Example}
\theoremstyle{programme}

\theoremstyle{property}

\theoremstyle{problem}

\title{ Generalized fuzzy rough sets based on fuzzy coverings}

\author
{Guangming Lang$^{a}$ \hspace{1cm} Qingguo Li$^{a}$
\thanks{Corresponding author.\quad Tel./fax: +86 731 8822855,
liqingguoli@yahoo.com.cn(Q. G. Li)
\newline\mbox{}\hspace{0.55cm}
E-mail address: langguangming1984@126.com(G. M. Lang),
lankun.guo@gmail.com(L. K. Guo).}\hspace{1cm}
Lankun Guo$^{b}$ \\
\small {$^{a}$ College of Mathematics and Econometrics, Hunan University}\\
\small {Changsha, Hunan 410082, P.R. China}\\
\small {$^{b}$ College of Information Science and Engineering, Hunan University}\\
\small {Changsha, Hunan 410082, P.R. China}}
\date{}

\begin{document}
\linespread{1.5}\selectfont
\newpage
$$ \text{\large\bf{Generalized fuzzy rough sets based on fuzzy coverings}}$$

Authors: Guangming Lang$^{a}$, Qingguo Li$^{a\ast}$ and Lankun
Guo$^{b}$

$^{\ast}$Corresponding author: liqingguoli@yahoo.com.cn

$^{a}$College of Mathematics and Econometrics, Hunan University,
Changsha, Hunan 410082, P.R. China

$^{b}$College of Information Science and Engineering, Hunan
University, Changsha, Hunan 410082, P.R. China\vskip.10in

{\bf Abstract.} This paper further studies the fuzzy rough sets
based on fuzzy coverings. We first present the notions of the lower
and upper approximation operators based on fuzzy coverings and
derive their basic properties. To facilitate the computation of
fuzzy coverings for fuzzy covering rough sets, the concepts of fuzzy
subcoverings, the reducible and intersectional elements, the union
and intersection operations are provided and their properties are
discussed in detail. Afterwards, we introduce the concepts of
consistent functions and fuzzy covering mappings and provide a basic
theoretical foundation for the communication between fuzzy covering
information systems. In addition, the notion of homomorphisms is
proposed to reveal the relationship between fuzzy covering
information systems. We show how large-scale fuzzy covering
information systems and dynamic fuzzy covering information systems
can be converted into small-scale ones by means of homomorphisms.
Finally, an illustrative example is employed to show that the
attribute reduction can be simplified significantly by our proposed
approach.

\maketitle \baselineskip=17pt
\begin{center}
\begin{quote}
\linespread{1.5}\selectfont{{\bf Abstract.} This paper further
studies the fuzzy rough sets based on fuzzy coverings. We first
present the notions of the lower and upper approximation operators
based on fuzzy coverings and derive their basic properties. To
facilitate the computation of fuzzy coverings for fuzzy covering
rough sets, the concepts of fuzzy subcoverings, the reducible and
intersectional elements, the union and intersection operations are
provided and their properties are discussed in detail. Afterwards,
we introduce the concepts of consistent functions and fuzzy covering
mappings and provide a basic theoretical foundation for the
communication between fuzzy covering information systems. In
addition, the notion of homomorphisms is proposed to reveal the
relationship between fuzzy covering information systems. We show how
large-scale fuzzy covering information systems and dynamic fuzzy
covering information systems can be converted into small-scale ones
by means of homomorphisms. Finally, an illustrative example is
employed to show that the attribute reduction can be simplified
significantly by our proposed approach.

\\}
\end{quote}
\end{center}
\renewcommand{\thesection}{\arabic{section}}

\linespread{1.5}\selectfont
\section{Introduction}
Rough set theory, originally constructed on the basis of an
equivalence relation, was proposed by Pawlak\cite{Pawlak1} for
solving inexact or uncertain problems. But the condition of the
equivalence relation is so restrictive that the applications of
rough sets are limited in many practical problems. To deal with more
complex data sets, many researchers have derived a large number of
generalized models by replacing the equivalence relation with a few
mathematical concepts such as fuzzy
relations\cite{Dubois1,Dubois2,Radzikowska1,Radzikowska2,Torabi,Korvin,Beaubouef}
and coverings\cite{Chen1,Deng1,Feng1,Ge1,Li1,
Pomykala1,Shi1,Tsang1,Yang1,Yao2,Zakowski1,
Zhang1,Zhang2,Zhu1,Zhu2,Zhu3,Zhu5,Zhu6} of the universe of
discourse.

Recently, the theory of fuzzy rough sets has become a rapidly
developing research area and got a lot of attention. For example,
Dubois et al.\cite{Dubois1,Dubois2} initially provided the rough
fuzzy sets and the fuzzy rough sets. Then Radzikowska et
al.\cite{Radzikowska1,Radzikowska2} defined the fuzzy rough sets
(respectively, the L-fuzzy rough sets) based on fuzzy similarity
relations (respectively, residuated lattices). Afterwards, many
researchers\cite{Li1,Zhang1,Deng1,Feng1} investigated fuzzy rough
sets based on fuzzy coverings. In practice, we need to compute the
approximations of fuzzy sets in fuzzy covering approximation spaces.
But the classical approximation operators based on coverings are
incapable of computing the approximations of fuzzy sets in the fuzzy
covering approximation space. It motivates us to extend
approximation operators of covering approximation spaces for fuzzy
covering approximation spaces. In addition, there are a large number
of fuzzy coverings for the universal set in general. To facilitate
the computation of fuzzy coverings for fuzzy covering rough sets, it
is interesting to investigate the relationship among the elements of
a fuzzy covering and operations on fuzzy coverings.

Meanwhile, many
researches\cite{Grzymala-Busse1,Grzymala-Busse2,Grzymala-Busse3,Li3,
Wang2,Wang3,Wang4,Wang5,Wang6,Gong1,Zhu7,Zhu8,Zhu9} have been
conducted on homomorphisms between information systems with the aim
of attribute reductions. For instance,
Grzymala-Busse\cite{Grzymala-Busse1,Grzymala-Busse2,Grzymala-Busse3}
initially introduced the concept of information system homomorphisms
and investigated its basic properties. Then Li et al.\cite{Li3}
discussed invariant characters of information systems under some
homomorphisms. Afterwards, Wang et al.\cite{Wang2} found that a
complex massive covering information system could be compressed into
a relatively small-scale one under the condition of a homomorphism,
and their attribute reductions are equivalent to each other.
Actually, we often deal with attribute reductions of large-scale
fuzzy covering information systems in practical situations, and the
work of Wang et al. mentioned above inspires that the attribute
reduction of large-scale fuzzy covering information systems may be
conducted by means of homomorphisms. But so far we have not seen any
work on homomorphisms between fuzzy covering information systems.
Additionally, the fuzzy covering information system varies with time
due to the dynamic characteristics of data collection, and the
non-incremental approach to compressing the dynamic fuzzy covering
information system is often very costly or even intractable. For
this issue, we attempt to apply an incremental updating scheme to
maintain the compression dynamically and avoid unnecessary
computations by utilizing the compression of the original system.

The purpose of this paper is to investigate further fuzzy coverings
based rough sets. First, we present the notions of the lower and
upper approximation operators based on fuzzy coverings by extending
Zhu's model\cite{Zhu3}, and examine their basic properties.
Particularly, we find that the upper approximation based on
neighborhoods can not be represented without using the neighborhoods
as the classical covering approximation space\cite{Zhu3} in the
fuzzy approximation space. Second, we propose the concepts of fuzzy
subcoverings, reducible and intersectional elements, union and
intersection operations and investigate their basic properties in
detail. Third, the theoretical foundation is established for the
communication between fuzzy covering information systems.
Concretely, we construct a consistent function by combining the
fuzzy covering, proposed by Deng et al. \cite{Deng1}, with the
approach in \cite{Wang2}, and explore its main properties known from
the consistent function for the classical covering approximation
space in \cite{Wang2}. We also provide the concepts of fuzzy
covering mappings and study their basic properties in detail.
Fourth, the notion of homomorphisms between fuzzy covering
information systems is introduced for attribute reductions. We find
that a large-scale fuzzy covering information system can be
compressed into a relatively small-scale one, and attribute
reductions of the original system and image system are equivalent to
each other under the condition of a homomorphism. In addition, we
give the algorithm to construct attribute reducts and employ an
example to illustrate the efficiency of our approach for attribute
reductions of fuzzy covering information systems. We also discuss
how to compress the dynamic fuzzy covering information system.

The rest of this paper is organized as follows: Section 2 briefly
reviews the basic concepts related to the covering information
systems and fuzzy covering information systems. In Section 3, we put
forward some concepts such as the neighborhood operators, the
approximation operators and reducible elements for fuzzy covering
approximation spaces, and investigate their basic properties.
Section 4 is devoted to introducing the concept of consistent
functions which provides the theoretical foundation for the
communication between fuzzy covering information systems. In Section
5, we present the notion of homomorphisms between fuzzy covering
information systems and discuss its basic properties.  We also
investigate data compressions of fuzzy covering information systems
and dynamic fuzzy covering information systems. An example is given
to illustrate that how to conduct attribute reductions of the fuzzy
covering information system by means of homomorphisms. We conclude
the paper and set further research directions in Section 6.

\section{Preliminaries}

In this section, we briefly recall some basic concepts related to
the covering information system and fuzzy covering information
system. Three examples are given to illustrate two types of covering
information systems.

\begin{definition} \cite{Bonikowski1} Let $U$ be a non-empty set $($the universe of
discourse$)$. A non-empty sub-family $\mathscr{C} \subseteq
\mathscr{P} (U)$ is called a covering of $U$ if

$(1)$ every element in $\mathscr{C} $ is non-empty;

$(2)$ $\bigcup \{C |C \in \mathscr{C} \}=U$, where $\mathscr{P} (U)$
is the powerset of $U$.
\end{definition}

It is clear that the concept of a covering is an extension of the
notion of a partition. In what follows, $(U, \mathscr{C}) $ is
called a classical covering approximation space.

To investigate further coverings based rough sets, Chen et al.
proposed the following concepts on coverings.

\begin{definition} \cite{Chen1} Let $\mathscr{C}$=$\{C_{1} , C_{2},
..., C_{N} \}$ be a covering of $U$, $C_{x} $=$\bigcap\{C_{i} | x\in
C_{i}  \text{ and } C_{i} \in \mathscr{C} \}$ for any $x\in U$, and
$Cov(\mathscr{C} )$=$\{C_{x} | x\in U\}$. Then $Cov(\mathscr{C} )$
is called the induced covering of $\mathscr{C} $.
\end{definition}

Suppose $c$ is an attribute, the domain of $c$ is $\{c_{1},
c_{2},...,c_{N}\}$, $C_{i}$ means the set of objects in $U$ taking a
certain attribute value $c_{i}$, and $C_{x}=C_{i}\cap C_{j}$, it
implies that the possible value of $x$ regarding the attribute $c$
is $c_{i}$ or $c_{j}$, and $C_{x}$ is the minimal set containing $x$
in $Cov(\mathscr{C})$.

\begin{definition} \cite{Chen1} Let $\Delta $=$\{\mathscr{C}_{1},
\mathscr{C}_{2} ,..., \mathscr{C}_{m} \}$ be a family of coverings
of $U$, $\Delta_{x} $=$\bigcap\{C_{ix} | C_{ix} \in \mathscr{C}_{i}
$, $1 \leq i\leq m\}$ for any $x\in U$, and $Cov(\Delta
)$=$\{\Delta_{x} | x\in U\}$. Then $Cov(\Delta )$ is called the
induced covering of $\Delta $.
\end{definition}

That is to say, $\Delta_{x}$ is the intersection of all the elements
including $x$ of each $\mathscr{C}_{i}$, and it is the minimal set
including $x$ in $Cov(\Delta )$. Furthermore, $Cov(\Delta )$ is a
partition if every covering in $\Delta $ is a partition. In what
follows, $(U, \Delta )$ is called a covering information system. To
illustrate how covering information systems are constructed, we
present two examples which have different application backgrounds.

\begin{example}\begin{upshape}
Let $U=\{x_{1},x_{2},x_{3},x_{4},x_{5},x_{6},x_{7},x_{8}\}$ be eight
houses, $C=\{price, color\}$ the attribute set, the domains of
$price$ and $color$ are $\{high, middle, low\}$ and $\{good, bad\}$,
respectively. To evaluate these houses, specialists  $A$ and $B$ are
employed and their evaluation reports are shown as follows:
\begin{eqnarray*}
high_{A}&=&\{x_{1}, x_{4}, x_{5}, x_{7}\}, middle_{A}=\{x_{2},
x_{8}\}, low_{A}=\{x_{3}, x_{6}\};\\
high_{B}&=&\{x_{1}, x_{2}, x_{4}, x_{7}, x_{8}\},
middle_{B}=\{x_{5}\}, low_{B}=\{x_{3}, x_{6}\};\\
good_{A}&=&\{x_{1}, x_{2}, x_{3}, x_{6}\}, bad_{A}=\{x_{4}, x_{5},
x_{7}, x_{8}\};\\
good_{B}&=&\{x_{1}, x_{2}, x_{3}, x_{5}\}, bad_{B}=\{x_{4}, x_{6},
x_{7}, x_{8}\},\end{eqnarray*} where $high_{A}$ denotes the houses
belonging to high price by the specialist $A$. The meanings of other
symbols are similar. Since their evaluations are of equal
importance, we should consider all their advice. Consequently, we
obtain the following results:
\begin{eqnarray*}
high_{A\vee B}&=&high_{A}\cup high_{B}=\{x_{1}, x_{2}, x_{4}, x_{5},
x_{7}, x_{8}\};\\  middle_{A\vee B}&=&middle_{A}\cup
middle_{B}=\{x_{2}, x_{5}, x_{8}\};\\ low_{A\vee B}&=&low_{A}\cup
low_{B}=\{x_{3}, x_{6}\};\\ good_{A\vee B}&=&good_{A}\cup
good_{B}=\{x_{1}, x_{2}, x_{3}, x_{5}, x_{6}\};\\ bad_{A\vee
B}&=&bad_{A}\cup bad_{B}=\{x_{4}, x_{5}, x_{6}, x_{7},
x_{8}\}.\end{eqnarray*}

Based on the above statement, we derive a covering information
system $(U, \Delta)$,  where $\Delta=\{\mathscr{C}_{price},$ $
\mathscr{C}_{color}\}$, $\mathscr{C}_{price}=\{high_{A\vee B},
middle_{A\vee B}, low_{A\vee B}\}$ and
$\mathscr{C}_{color}=\{good_{A\vee B}, bad_{A\vee
B}\}$.\end{upshape}\end{example}

\begin{example}\begin{upshape}
Let Table 1 be an incomplete information system, where $\ast$ stands
for the lost value. According to the interpretation in
\cite{Grzymala-Busse3}, the lost value is considered to be similar
to any value in the domain of the corresponding attribute.
Consequently, we obtain three coverings of $U$ by the attribute set
as follows: $\mathscr{C}_{structure}=\{\{x_{1}, x_{2}, x_{4},
x_{6}\}, \{x_{2}, x_{3}, x_{5}, x_{6}\}\}$,
$\mathscr{C}_{color}=\{\{x_{1}, x_{2}, x_{5}\}, \{x_{3}, x_{4},
x_{5}, x_{6}\}\}$, $\mathscr{C}_{price}=\{\{x_{1}, x_{4}, x_{5},
x_{6}\}, \{x_{2}, x_{3}, x_{4}, x_{6}\}\}$. Hence, $(U, \Delta)$ is
a covering information system, where
$\Delta=\{\mathscr{C}_{structure}, \mathscr{C}_{color},$
$\mathscr{C}_{price}\}$.\end{upshape}
\end{example}

To conduct the communication between covering information systems,
Wang et al. provided the concept of consistent functions based on
coverings.

\begin{definition}\cite{Wang2}
Let $f$ be a mapping from $U_{1}$ to $U_{2}$,
$\mathscr{C}$=$\{C_{1}, C_{2},...,C_{N}\}$    a covering of $U_{1}$,
$C_{x} $=$\bigcap\{C_{i} | x\in C_{i}  \text{ and } C_{i} \in
\mathscr{C} \}$, and $[x]_{f}=\{y\in U_{1}| f(x)=f(y)\}$.  If
$[x]_{f}\subseteq C_{x}$ for any $x\in U_{1}$, then $f$ is called a
consistent function with respect to $\mathscr{C}$.
\end{definition}

Based on Definition 2.6, Wang et al. constructed a homomorphism
between a complex massive covering information system and a
relatively small-scale covering information system. It has also been
proved that their attribute reductions are equivalent to each other
under the condition of a homomorphism. Hence, the notion of the
consistent function provides the foundation for the communication
between covering information systems.

In order to deal with uncertainty and more complex problems,
Zadeh\cite{Zadeh1} proposed the theory of fuzzy sets by extending
the classical set theory. Let $U$ be a non-empty universe of
discourse, a fuzzy set of $U$ is a mapping $A: U\longrightarrow [0,
1]$. We denote by $\mathscr{F}(U)$ the set of all fuzzy sets of $U$.
For any $A, B\in \mathscr{F}(U)$, we say that $A\subseteq B$ if
$A(x)\leq B(x)$ for any $x\in U$. The union of $A$ and $B$, denoted
as $A\cup B$, is defined by $(A\cup B)(x)=A(x)\vee B(x)$ for any
$x\in U$, and the intersection of $A$ and $B$, denoted as $A\cap B$,
is defined by $(A\cap B)(x)=A(x)\wedge B(x)$ for any $x\in U$. The
complement of $A$, denoted as $-A$, is defined by $(-A)(x)=1-A(x)$
for any $x\in U$. Furthermore, a fuzzy relation on $U$ is a mapping
$R: U\times U\longrightarrow [0, 1]$. We denote by
$\mathscr{F}(U\times U)$ the set of all fuzzy relations on $U$.

In practical situations, there exist a lot of fuzzy information
systems as a generalization of crisp information systems, and the
investigations of fuzzy information systems have powerful prospects
in applications. To conduct the communication between fuzzy
information systems, Wang et al. proposed a consistent function with
respect to a fuzzy relation.

\begin{definition}\cite{Wang5}
Let $U_{1}$ and $U_{2}$ be two universes, $f$ a mapping from $U_{1}$
to $U_{2}$, $R\in \mathscr{F}(U_{1}\times U_{1})$, $[x]_{f}=\{y\in
U_{1}|f(x)=f(y)\}$, and $\{[x]_{f}|x\in U_{1}\}$ is a partition on
$U_{1}$. For any $x,y\in U_{1}$, if $R(u,v)=R(s,t)$ for any two
pairs $(u,v),(s,t)\in [x]_{f}\times[y]_{f}$, then $f$ is said to be
consistent with respect to $R$.
\end{definition}

Based on the consistent function, Wang et al. constructed a
homomorphism between a large-scale fuzzy information system and a
relatively small-scale fuzzy information system. It has been proved
that their attribute reductions are equivalent to each other under
the condition of a homomorphism. In this sense, the notion of the
consistent function provides an approach to studying the
communication between fuzzy information systems.

Recently, Deng et al.\cite{Deng1} proposed the concept of a fuzzy
covering.

\begin{definition} \cite{Deng1} A fuzzy covering of  $U$ is a collection
of fuzzy sets
$\mathscr{C}^{\ast}\subseteq \mathscr{F}(U)$  which satisfies

$(1)$ every fuzzy set $C^{\ast}\in \mathscr{C}^{\ast}$ is non-null,
i.e., $C^{\ast}\neq\emptyset$;

$(2)$ $ \forall x\in U, \bigvee_{C^{\ast}\in
\mathscr{C}^{\ast}}C^\ast(x)>0$.
\end{definition}

Unless stated otherwise, $U$ is a finite universe, and
$\mathscr{C}^{\ast}$ consists of finite number of sets in this work.
In what follows, $(U, \mathscr{C}^{\ast}) $ is called a fuzzy
covering approximation space, and $(U, \Delta^{\ast} )$ is called a
fuzzy covering information system, where
$\Delta^{\ast}=\{\mathscr{C}_{i}^{\ast}|1\leq i\leq m\}$. Throughout
the paper, we denote the set of all fuzzy coverings of $U$ as $C(U)$
for simplicity.

In the following, we employ an example to illustrate the fuzzy
covering information system.

\begin{example}\begin{upshape}
Let $U=\{x_{1},x_{2},x_{3},x_{4},x_{5},x_{6},x_{7},x_{8}\}$ be eight
houses, $C=\{price, color\}$ the attribute set, the domains of
$price$ and $color$ are $\{high, middle, low\}$ and $\{good, bad\}$,
respectively. To evaluate these houses, specialists $A$ and $B$ are
employed and their evaluation reports are shown as follows:
\begin{eqnarray*}high^{\ast}_{A}&=&\frac{1}{x_{1}}+\frac{0.7}{x_{2}}+\frac{0}{x_{3}}
+\frac{1}{x_{4}}+\frac{1}{x_{5}}
+\frac{0}{x_{6}}+\frac{1}{x_{7}}+\frac{0.65}{x_{8}};\\
middle^{\ast}_{A}&=&\frac{0.6}{x_{1}}+\frac{1}{x_{2}}+\frac{0.4}{x_{3}}
+\frac{0.4}{x_{4}}+\frac{0.45}{x_{5}}
+\frac{0.5}{x_{6}}+\frac{0.5}{x_{7}}+\frac{1}{x_{8}};\\
low^{\ast}_{A}&=&\frac{0}{x_{1}}+\frac{0.5}{x_{2}}+\frac{1}{x_{3}}
+\frac{0}{x_{4}}+\frac{0.5}{x_{5}}
+\frac{1}{x_{6}}+\frac{0}{x_{7}}+\frac{0.5}{x_{8}};\\
good^{\ast}_{A}&=&\frac{1}{x_{1}}+\frac{1}{x_{2}}+\frac{1}{x_{3}}
+\frac{0.5}{x_{4}}+\frac{0.6}{x_{5}}
+\frac{1}{x_{6}}+\frac{0}{x_{7}}+\frac{0}{x_{8}};\\
bad^{\ast}_{A}&=&\frac{0}{x_{1}}+\frac{0.4}{x_{2}}+\frac{0}{x_{3}}
+\frac{1}{x_{4}}+\frac{1}{x_{5}}
+\frac{0.2}{x_{6}}+\frac{1}{x_{7}}+\frac{1}{x_{8}};\\
high_{B}&=&\frac{0.9}{x_{1}}+\frac{0.7}{x_{2}}+\frac{0}{x_{3}}
+\frac{1}{x_{4}}+\frac{1}{x_{5}}
+\frac{0}{x_{6}}+\frac{1}{x_{7}}+\frac{0.8}{x_{8}};\\
middle^{\ast}_{B}&=&\frac{0.6}{x_{1}}+\frac{1}{x_{2}}+\frac{0.4}{x_{3}}
+\frac{0.4}{x_{4}}+\frac{0.45}{x_{5}}
+\frac{0.7}{x_{6}}+\frac{0.5}{x_{7}}+\frac{1}{x_{8}};\\
low^{\ast}_{B}&=&\frac{0}{x_{1}}+\frac{0.5}{x_{2}}+\frac{1}{x_{3}}
+\frac{0}{x_{4}}+\frac{0.5}{x_{5}}
+\frac{0.9}{x_{6}}+\frac{0}{x_{7}}+\frac{0.5}{x_{8}};\\
good^{\ast}_{B}&=&\frac{0.8}{x_{1}}+\frac{1}{x_{2}}+\frac{0.9}{x_{3}}
+\frac{0.5}{x_{4}}+\frac{0.6}{x_{5}}
+\frac{1}{x_{6}}+\frac{0}{x_{7}}+\frac{0}{x_{8}};\\
bad^{\ast}_{B}&=&\frac{0}{x_{1}}+\frac{0.4}{x_{2}}+\frac{0.4}{x_{3}}
+\frac{1}{x_{4}}+\frac{1}{x_{5}}
+\frac{0.2}{x_{6}}+\frac{0.9}{x_{7}}+\frac{1}{x_{8}},\end{eqnarray*}
where $high_{A}$ is the membership degree of each house belonging to
the high price by the specialist $A$. The meanings of the other
symbols are similar. Based on the above results, we obtain a fuzzy
covering information system $(U, \Delta^{\ast})$, where
$\Delta^{\ast}=\{\mathscr{C}^{\ast}_{price},
\mathscr{C}^{\ast}_{color}\}$,
$\mathscr{C}^{\ast}_{price}=\{C_{high}, C_{middle}, C_{low}\}$ and
$\mathscr{C}^{\ast}_{color}=\{C_{good}, C_{bad}\}$.
\begin{eqnarray*}
C_{high}&=&high^{\ast}_{A}\cup
high^{\ast}_{B}=\frac{1}{x_{1}}+\frac{0.7}{x_{2}}+\frac{0}{x_{3}}
+\frac{1}{x_{4}}+\frac{1}{x_{5}}
+\frac{0}{x_{6}}+\frac{1}{x_{7}}+\frac{0.8}{x_{8}};\\
C_{middle}&=&middle^{\ast}_{A}\cup
middle^{\ast}_{B}=\frac{0.6}{x_{1}}+\frac{1}{x_{2}}+\frac{0.4}{x_{3}}
+\frac{0.4}{x_{4}}+\frac{0.45}{x_{5}}
+\frac{0.7}{x_{6}}+\frac{0.5}{x_{7}}+\frac{1}{x_{8}};\\
C_{low}&=&low^{\ast}_{A}\cup
low^{\ast}_{B}=\frac{0}{x_{1}}+\frac{0.5}{x_{2}}+\frac{1}{x_{3}}
+\frac{0}{x_{4}}+\frac{0.5}{x_{5}}
+\frac{1}{x_{6}}+\frac{0}{x_{7}}+\frac{0.5}{x_{8}};\\
C_{good}&=&good^{\ast}_{A}\cup
good^{\ast}_{B}=\frac{1}{x_{1}}+\frac{1}{x_{2}}+\frac{1}{x_{3}}
+\frac{0.5}{x_{4}}+\frac{0.6}{x_{5}}
+\frac{1}{x_{6}}+\frac{0}{x_{7}}+\frac{0}{x_{8}};\\
C_{bad}&=&bad^{\ast}_{A}\cup
bad^{\ast}_{B}=\frac{0}{x_{1}}+\frac{0.4}{x_{2}}+\frac{0.4}{x_{3}}
+\frac{1}{x_{4}}+\frac{1}{x_{5}}
+\frac{0.2}{x_{6}}+\frac{1}{x_{7}}+\frac{1}{x_{8}}.\end{eqnarray*}
\end{upshape}\end{example}

It is obvious that we can construct a fuzzy covering of the universe
with an attribute. Since the fuzzy covering rough set theory is
effective to handle uncertain information, the investigation of this
theory becomes an important task in rough set theory.

\section{The basic properties of the fuzzy covering approximation space}

In this section, we introduce the concepts of neighborhoods, the
lower and upper approximation operators to facilitate the
computation of fuzzy sets for fuzzy covering approximation spaces.
Then we propose the concepts of fuzzy subcoverings, irreducible and
reducible elements, non-intersectional and intersectional elements
of fuzzy coverings. Afterwards, the union and intersection
operations on two fuzzy coverings are provided. We also construct
two roughness measures and employ several examples to illustrate the
proposed notions.
\subsection{The lower and upper approximation operations}

Before introducing approximation operators, we present the concepts
of neighborhoods and induced fuzzy coverings based on fuzzy
coverings.

\begin{definition}
Let $(U, \mathscr{C}^{\ast})$ be a fuzzy covering approximation
space, and $x\in U$. Then
$C^{\ast}_{\mathscr{C}^{\ast}x}=\bigcap\{C^{\ast}|C^{\ast}(x)$ $>0
\text{ and } C^{\ast}\in \mathscr{C}^{\ast}\}$ is called the
neighborhood of $x$ concerning $\mathscr{C}^{\ast}$.
\end{definition}

We notice that $C_{x}^{\ast}$ is the intersection of all fuzzy
subsets whose membership degrees of $x\in U$ are not zeroes. Assume
that $C_{1}^{\ast},C_{2}^{\ast}\in\mathscr{C}^{\ast}$,
$C_{1}^{\ast}(x)>0, C_{2}^{\ast}(x)>0$, and
$C^{\ast}_{x}=C^{\ast}_{1}\cap C^{\ast}_{2}$ for $x\in U$, it
implies that the membership degree of $x$ in $C^{\ast}_{x}$ is
$min\{C_{1}^{\ast}(x), C_{2}^{\ast}(x)\}$. In addition, we observe
that the classical neighborhood of a point $C_{x}=\bigcap\{C|x\in
C\in \mathscr{C}\}$ is the same as that in Definition 3.1 if the
membership degree for any $x\in U$ has its value only from the set
$\{0, 1\}$, where $\mathscr{C}$ is a covering of $U$. For
convenience, we denote $C^{\ast}$, $\mathscr{C}^{\ast}$,
$C^{\ast}_{\mathscr{C}_{i}x}$  and $C_{\mathscr{C}x}$ as $C$,
$\mathscr{C}$, $C_{ix}$ and $C_{x}$, respectively.

We present the properties of the neighborhood operator below.

\begin{proposition}
Let $(U, \mathscr{C})$ be a fuzzy covering approximation space, and
$x, y\in U$. If $C_{x}(y)>0$, then $C_{y}\subseteq C_{x}.$
\end{proposition}
\noindent\textbf{Proof.} Assume that
$\mathscr{A}=\{C|C\in\mathscr{C}, C(x)>0\}$ and
$\mathscr{B}=\{C'|C'\in\mathscr{C}, C'(y)>0\}$. Since $C_{x}(y)>0$,
it follows that $C(y)>0$ for any $C\in \mathscr{A}$. Consequently,
$C\in \mathscr{B}$. It implies that $\{C|C\in\mathscr{C},
C(x)>0\}\subseteq\{C'|C'\in\mathscr{C}, C'(y)>0\}.$ Therefore,
$C_{y}\subseteq C_{x}.$        $\Box$

\begin{proposition}
Let $(U, \mathscr{C})$ be a fuzzy covering approximation space, and
$x, y\in U$. If $C_{x}(y)>0$ and $C_{y}(x)>0$, then $C_{y}= C_{x}.$
\end{proposition}
\noindent\textbf{Proof.} Straightforward from Proposition 3.2.
$\Box$

Based on Definition 3.1, we present the concept of a fuzzy covering
induced by the original fuzzy covering.

\begin{definition} Let $\mathscr{C}$=$\{C_{1}, C_{2},..., C_{N}\}$ be a
fuzzy  covering of $U$,  $C_{x}$=$\bigcap\{C_{i}| C_{i}(x)>0 \text{
and } C_{i}\in \mathscr{C}\}$ for any $x\in U$,  and
$Cov(\mathscr{C})$=$\{C_{x}| x\in U\}$. Then $Cov(\mathscr{C})$ is
called the induced fuzzy covering of $\mathscr{C}$.
\end{definition}

It is clear that $C_{x}$ has the minimal membership degree of $x$ in
$Cov(\mathscr{C})$, and each element of $Cov(\mathscr{C})$ can not
be represented as the union of other elements of $Cov(\mathscr{C})$.
In other words, $C_{x}$ is the minimal set containing $x$ in
$Cov(\mathscr{C})$. Furthermore, $Cov(\mathscr{C})$ is a fuzzy
covering of $U$, and it is easy to prove that the concept presented
in Definition 2.2 is a special case of Definition 3.4 when the
values of membership degree are taken from the set $\{0, 1\}$.

An example is employed to illustrate the induced fuzzy covering.

\begin{example}
\begin{upshape}
Let $U_{1}=\{x_{1},x_{2},x_{3},x_{4}\}$, and
$\mathscr{C}_{1}=\{C'_{1},C'_{2},C'_{3}\}$, where
$C'_{1}=\frac{1}{x_{1}}+\frac{0.5}{x_{2}}+\frac{1}{x_{3}}+\frac{0.5}{x_{4}}$,
$C'_{2}=\frac{0.5}{x_{1}}+\frac{0.6}{x_{2}}+\frac{0.5}{x_{3}}+\frac{0.6}{x_{4}}$,
and
$C'_{3}=\frac{0}{x_{1}}+\frac{0.5}{x_{2}}+\frac{0}{x_{3}}+\frac{0.5}{x_{4}}$.
By Definition 3.4, we obtain the induced fuzzy covering
$Cov(\mathscr{C}_{1})=\{C_{x_{i}}|i=1,2,3,4\}$, where
$C_{x_{1}}=C_{x_{3}}=\frac{0.5}{x_{1}}+\frac{0.5}{x_{2}}+\frac{0.5}{x_{3}}+\frac{0.5}{x_{4}}$,
and
$C_{x_{2}}=C_{x_{4}}=\frac{0}{x_{1}}+\frac{0.5}{x_{2}}+\frac{0}{x_{3}}+\frac{0.5}{x_{4}}$.
\end{upshape}
\end{example}

For convenience, we denote $C'_{i}$ as $C_{i}$ in the following
examples.

We also propose the notion of a fuzzy covering induced by a family
of fuzzy coverings.

\begin{definition} Let $\Delta$=$\{\mathscr{C}_{1}, \mathscr{C}_{2},...,
\mathscr{C}_{m}\}$ be a family of fuzzy  coverings of  $U$,
$\Delta_{x}$=$\bigcap\{C_{ix}| C_{ix}\in Cov(\mathscr{C}_{i})$,
$1\leq i\leq m\}$ for any $x\in U$, and $Cov(\Delta)$=$\{\Delta_{x}|
x\in U\}$. Then $Cov(\Delta)$ is called the induced  fuzzy covering
of $\Delta$.
\end{definition}

In other words, $\Delta_{x}$ is the intersection of all the elements
whose membership degrees of $x$ are not zeroes in each
$\mathscr{C}_{i}$, and it is the set whose membership degree of $x$
is the minimal in $Cov(\Delta )$. Furthermore, given $x, y\in U$, if
$\Delta_{x}(y)>0$, then $\Delta_{y}\subseteq \Delta_{x}$.
Consequently, $\Delta_{x}(y)>0$ and $\Delta_{y}(x)>0$ imply that
$\Delta_{x}=\Delta_{y}$. In addition, $Cov(\Delta)$ is a fuzzy
covering of $U$. Therefore, it is easy to verify that the notion
given in Definition 2.3 is a special case of Definition 3.6 when the
values of membership degree are taken from the set $\{0, 1\}$.

Next, we give an example to illustrate Definition 3.6.

\begin{example}\begin{upshape}
Let $U_{1}=\{x_{1},x_{2},x_{3},x_{4}\}$,
$\Delta=\{\mathscr{C}_{1},\mathscr{C}_{2},\mathscr{C}_{3}\}$,
$\mathscr{C}_{1}=\{C_{4},C_{5},C_{6}\}$,
$\mathscr{C}_{2}=\{C_{7},C_{8},C_{9}\}$, and
$\mathscr{C}_{3}=\{C_{10},C_{11},C_{12}\}$, where
$C_{4}=\frac{1}{x_{1}}+\frac{1}{x_{2}}+\frac{0.5}{x_{3}}+\frac{0.5}{x_{4}}$,
$C_{5}=\frac{0.5}{x_{1}}+\frac{0.5}{x_{2}}+\frac{0.6}{x_{3}}+\frac{0.6}{x_{4}}$,
$C_{6}=\frac{0}{x_{1}}+\frac{0}{x_{2}}+\frac{0.5}{x_{3}}+\frac{0.5}{x_{4}}$,
$C_{7}=\frac{0}{x_{1}}+\frac{0}{x_{2}}+\frac{1}{x_{3}}+\frac{1}{x_{4}}$
$C_{8}=\frac{1}{x_{1}}+\frac{1}{x_{2}}+\frac{0.7}{x_{3}}+\frac{0.7}{x_{4}}$,
$C_{9}=\frac{0.6}{x_{1}}+\frac{0.6}{x_{2}}+\frac{0.5}{x_{3}}+\frac{0.5}{x_{4}}$,
$C_{10}=\frac{1}{x_{1}}+\frac{1}{x_{2}}+\frac{1}{x_{3}}+\frac{1}{x_{4}}$
$C_{11}=\frac{0.5}{x_{1}}+\frac{0.5}{x_{2}}+\frac{1}{x_{3}}+\frac{1}{x_{4}}$,
and
$C_{12}=\frac{0.8}{x_{1}}+\frac{0.8}{x_{2}}+\frac{0.7}{x_{3}}+\frac{0.7}{x_{4}}$.
By Definition 3.6, we obtain that
$Cov(\Delta)=\{\Delta_{x_{i}}|i=1,2,3,4\}$, where
$\Delta_{x_{1}}=\Delta_{x_{2}}=\frac{0.5}{x_{1}}+\frac{0.5}{x_{2}}+\frac{0.5}{x_{3}}+\frac{0.5}{x_{4}}$,
and
$\Delta_{x_{3}}=\Delta_{x_{4}}=\frac{0}{x_{1}}+\frac{0}{x_{2}}+\frac{0.5}{x_{3}}+\frac{0.5}{x_{4}}$.
\end{upshape}\end{example}

In practice, the classical approximation operators based on
coverings are not fit for computing the approximations of fuzzy sets
in the fuzzy covering approximation space. To solve this issue, we
propose the concepts of the lower and upper approximation operators
based on fuzzy coverings by extending approximation operators in
\cite{Zhu3}.

\begin{definition}
Let $(U, \mathscr{C})$ be a fuzzy covering approximation space, and
$X\subseteq U$. Then the lower and upper approximations of X are
defined as
\begin{eqnarray*}
\underline{X}_{\mathscr{C}}&=&\bigcup\{C|C\subseteq X \text{ and }
C\in \mathscr{C}\};\\
\overline{X}_{\mathscr{C}}&=&\left(\bigcup\{C_{x}|X(x)>0 \text{ and
} \underline{X}_{\mathscr{C}}(x)=0, x\in U\}\right)\cup
\underline{X}_{\mathscr{C}}.\end{eqnarray*}
\end{definition}

The physical meaning of the lower and upper approximations of $X$ is
that we can approximate $X$ by $\underline{X}_{\mathscr{C}}$ and
$\overline{X}_{\mathscr{C}}$. Particularly, if
$\overline{X}_{\mathscr{C}}=\underline{X}_{\mathscr{C}}=X$, then $X$
can be understood as a definable set. Otherwise, $X$ is undefinable.
It is clear that the lower and upper approximation operations are
the same as those\cite{Zhu3} in the classical covering approximation
space if $\mathscr{C}$ is a covering of $U$. In this sense, the
notions given in Definition 3.8 are generalizations of the classical
ones into the fuzzy setting. In the following, we investigate their
basic properties in detail.

\begin{proposition}
Let $(U, \mathscr{C})$ be a fuzzy covering approximation space, and
$X, Y\subseteq U$. Then

$(1)$ $\overline{\emptyset}_{\mathscr{C}}= \emptyset$,
$\underline{\emptyset}_{\mathscr{C}}= \emptyset$;

$(2)$ $\underline{U}_{\mathscr{C}}\subseteq U$,
$\overline{U}_{\mathscr{C}}\subseteq U$;

$(3)$ $\underline{X}_{\mathscr{C}}\subseteq
\overline{X}_{\mathscr{C}}$, $\underline{X}_{\mathscr{C}}\subseteq
X$;

$(4)$ $\overline{X}_{\mathscr{C}}\cup
\overline{Y}_{\mathscr{C}}\subseteq \overline{(X\cup
Y)}_{\mathscr{C}}$;

$(5)$ $X\subseteq
Y\Longrightarrow\underline{X}_{\mathscr{C}}\subseteq
\underline{Y}_{\mathscr{C}}, \overline{X}_{\mathscr{C}}\subseteq
\overline{Y}_{\mathscr{C}}$;

$(6)$ $\forall C\in \mathscr{C}$, $\underline{C}=C, \overline{C}=C;$

$(7)$ $\underline{(\underline{X}_{\mathscr{C}})}_{\mathscr{C}}
=\underline{X}_{\mathscr{C}}$, $\overline{X}_{\mathscr{C}}=
\overline{(\overline{X}_{\mathscr{C}})}_{\mathscr{C}}$;

$(8)$ $\overline{(\underline{X}_{\mathscr{C}})}_{\mathscr{C}}
=\underline{X}_{\mathscr{C}}$,
$\underline{(\overline{X}_{\mathscr{C}})}_{\mathscr{C}}
\subseteq\overline{X}_{\mathscr{C}}$.
\end{proposition}

\noindent\textbf{Proof.} Straightforward from Definition 3.8. $\Box$

\begin{proposition}
The following properties do not hold generally in the fuzzy covering
approximation space:

$(1)$ $\underline{(X\cap
Y)}_{\mathscr{C}}=\underline{X}_{\mathscr{C}}\cap
\underline{Y}_{\mathscr{C}}$;

$(2)$
$\underline{(-X)}_{\mathscr{C}}=-(\overline{X}_{\mathscr{C}})$;

$(3)$
$\overline{(-X)}_{\mathscr{C}}=-(\underline{X}_{\mathscr{C}})$;

$(4)$
$\underline{(-\underline{X}_{\mathscr{C}})}_{\mathscr{C}}=-(\underline{X}_{\mathscr{C}})$;

$(5)$
$\overline{(-\overline{X}_{\mathscr{C}})}_{\mathscr{C}}=-(\overline{X}_{\mathscr{C}})$;

$(6)$ $\underline{U}_{\mathscr{C}}= U$, $\overline{U}_{\mathscr{C}}=
U$;

$(7)$ $\overline{(X\cup
Y)}_{\mathscr{C}}\subseteq\overline{X}_{\mathscr{C}}\cup
\overline{Y}_{\mathscr{C}}$;

$(8)$ $X\subseteq \overline{X}_{\mathscr{C}}$.
\end{proposition}

Example 2 in \cite{Zhu1} can illustrate that Proposition 3.10(1-5)
does not hold generally in the fuzzy covering approximation space.
Specially, we obtain that $\underline{U}_{\mathscr{C}}= U$ and
$X\subseteq \overline{X}_{\mathscr{C}}$ do not necessarily hold for
any $X\subseteq U$. Consequently, the lower and upper approximation
operations are not interior and closure operators, respectively, in
the fuzzy covering approximation space.

We employ an example to illustrate that Proposition 3.10(6-8) does
not hold generally.

\begin{example}\begin{upshape}
Let $U=\{x_{1}, x_{2}, x_{3}, x_{4}\}$, $\mathscr{C}=\{C_{13},
C_{14}, C_{15}, $ $C_{16}\}$, where
$C_{13}=\frac{0.3}{x_{1}}+\frac{0}{x_{2}}+\frac{0}{x_{3}}+\frac{0}{x_{4}},$
$C_{14}=\frac{0}{x_{1}}+\frac{0}{x_{2}}+\frac{0.5}{x_{3}}+\frac{0.5}{x_{4}},$
$C_{15}=\frac{0.3}{x_{1}}+\frac{0}{x_{2}}+\frac{0}{x_{3}}+\frac{0.4}{x_{4}}$,
and
$C_{16}=\frac{0}{x_{1}}+\frac{0.4}{x_{2}}+\frac{0.5}{x_{3}}+\frac{0}{x_{4}}.$
By Definition 3.8, it follows that
$\overline{U}_{\mathscr{C}}=\underline{U}_{\mathscr{C}}=\frac{0.3}{x_{1}}+\frac{0.4}{x_{2}}+
\frac{0.5}{x_{3}}+\frac{0.5}{x_{4}}\neq U$. For
$X=\frac{0.4}{x_{1}}+\frac{0}{x_{2}}+\frac{0.1}{x_{3}}+\frac{0.5}{x_{4}}$
and
$Y=\frac{0}{x_{1}}+\frac{0.5}{x_{2}}+\frac{0.5}{x_{3}}+\frac{0}{x_{4}}$,
we have that
$\overline{X}_{\mathscr{C}}=\frac{0.3}{x_{1}}+\frac{0}{x_{2}}+\frac{0.5}{x_{3}}+\frac{0.4}{x_{4}}$,
$\overline{Y}_{\mathscr{C}}=\frac{0}{x_{1}}+\frac{0.4}{x_{2}}+\frac{0.5}{x_{3}}+\frac{0}{x_{4}}$
and $\overline{(X\cup
Y)}_{\mathscr{C}}=\frac{0.3}{x_{1}}+\frac{0.4}{x_{2}}+\frac{0.5}{x_{3}}+\frac{0.5}{x_{4}}$.
Consequently, $\overline{(X\cup
Y)}_{\mathscr{C}}\neq\overline{X}_{\mathscr{C}}\cup\overline{Y}_{\mathscr{C}}$
and $X\nsubseteq \overline{X}_{\mathscr{C}}$.
\end{upshape}\end{example}

Some relationships among $\underline{X}_{\mathscr{C}}$,
$\overline{X}_{\mathscr{C}}$ and $X$ are explored in the following.

\begin{proposition} Let $(U, \mathscr{C})$ be a fuzzy covering approximation space, and
$X\subseteq U$.

$(1)$ If $\underline{X}_{\mathscr{C}}=X$, then
$\overline{X}_{\mathscr{C}}=\underline{X}_{\mathscr{C}}$;

$(2)$ If $\underline{X}_{\mathscr{C}}=X$, then
$\overline{X}_{\mathscr{C}}=X$;

$(3)$ $\underline{X}_{\mathscr{C}}=X$ if and only if $X$ is a union
of elements in $\mathscr{C}$;

$(4)$ If $X$ is a union of elements in $\mathscr{C}$, then
$\overline{X}_{\mathscr{C}}=X$.

\end{proposition}

Next, an example is given to illustrate that the converses of
Proposition 3.12(1), (2) and (4) do not hold generally.

\begin{example}\begin{upshape}
Let $U=\{x_{1}, x_{2}, x_{3}, x_{4}\}$, $\mathscr{C}=\{C_{17},
C_{18}, C_{19}, $ $C_{20}\}$, where
$C_{17}=\frac{0.2}{x_{1}}+\frac{0.1}{x_{2}}+\frac{0.1}{x_{3}}+\frac{0.1}{x_{4}},$
$C_{18}=\frac{0.1}{x_{1}}+\frac{0.2}{x_{2}}+\frac{0.1}{x_{3}}+\frac{0.1}{x_{4}},$
$C_{19}=\frac{0.1}{x_{1}}+\frac{0.1}{x_{2}}+\frac{0.2}{x_{3}}+\frac{0.1}{x_{4}}$,
and
$C_{20}=\frac{0.1}{x_{1}}+\frac{0.1}{x_{2}}+\frac{0.1}{x_{3}}+\frac{0.2}{x_{4}}.$
For
$X=\frac{0.5}{x_{1}}+\frac{0.5}{x_{2}}+\frac{0.5}{x_{3}}+\frac{0.5}{x_{4}}$,
it follows that $\underline{X}_{\mathscr{C}}=\frac{0.2}{x_{1}}
+\frac{0.2}{x_{2}}+\frac{0.2}{x_{3}}+\frac{0.2}{x_{4}}
=\overline{X}_{\mathscr{C}}$. But $X$ is not a union of some subsets
in the fuzzy covering $\mathscr{C}$. For
$Y=\frac{0.1}{x_{1}}+\frac{0.1}{x_{2}}+\frac{0.1}{x_{3}}+\frac{0.1}{x_{4}}$,
according to Definition 3.1, it follows that $C_{x_{1}} =C_{x_{2}}
=C_{x_{3}} =C_{x_{4}}=\frac{0.1}{x_{1}}+\frac{0.1}
{x_{2}}+\frac{0.1}{x_{3}}+\frac{0.1}{x_{4}}$. Then we have that
$\overline{Y}=Y$, but $Y$ is not a union of some elements of
$\mathscr{C}$. \end{upshape}\end{example}

From Proposition 3.10, we see that $\underline{X}_{\mathscr{C}}\cap
\underline{Y}_{\mathscr{C}}=\underline{(X\cap Y)}_{\mathscr{C}}$
does not hold generally for any $X,Y\subseteq U$ in the fuzzy
covering approximation space. But if
$\underline{X}_{\mathscr{C}}\cap
\underline{Y}_{\mathscr{C}}=\underline{(X\cap Y)}_{\mathscr{C}}$ for
any $X,Y\subseteq U$, then we can obtain the following results.

\begin{proposition}
If $\underline{X}_{\mathscr{C}}\cap
\underline{Y}_{\mathscr{C}}=\underline{(X\cap Y)}_{\mathscr{C}}$ for
any $X,Y\subseteq U$, then $C_{1}\cap C_{2}=\emptyset$ or $C_{1}\cap
C_{2}$ is a union of elements of $\mathscr{C}$ for any $C_{1},
C_{2}\in \mathscr{C}$.
\end{proposition}

\noindent\textbf{Proof.} Taking any $C_{1}, C_{2}\in \mathscr{C},$
it follows that $\underline{C_{1}}_{\mathscr{C}}\cap
\underline{C_{2}}_{\mathscr{C}}=\underline{(C_{1}\cap
C_{2})}_{\mathscr{C}}=C_{1}\cap C_{2}.$ By Proposition 3.12, we have
that $C_{1}\cap C_{2}=\emptyset$ or $C_{1}\cap C_{2}$ is a union of
elements of $\mathscr{C}$ for any $C_{1}, C_{2}\in \mathscr{C}$.
       $\Box$

This proposition shows that the intersection of two elementary
elements in a fuzzy covering $\mathscr{C}$ can be represented as a
union of elements of $\mathscr{C}$ if
$\underline{X}_{\mathscr{C}}\cap
\underline{Y}_{\mathscr{C}}=\underline{(X\cap Y)}_{\mathscr{C}}$ for
any $X,Y\subseteq U$.

It is clear that $\underline{X}_{\mathscr{C}}\subseteq
\underline{X}_{Cov(\mathscr{C})}$ and $\overline{X}_{\mathscr{C}}=
\overline{X}_{Cov(\mathscr{C})}$ for any $X\subseteq U$ in the
classical covering approximation space $(U, \mathscr{C})$. But they
do not necessarily hold in the fuzzy covering approximation space.
To illustrate this point, we employ the following example.

\begin{example}\begin{upshape}
Let $U=\{x_{1}, x_{2}, x_{3}, x_{4}\}$, $\mathscr{C}=\{C_{21},
C_{22},$ $ C_{23}\}$, where
$C_{21}=\frac{0.2}{x_{1}}+\frac{0.4}{x_{2}}+\frac{0.5}{x_{3}}+\frac{0}{x_{4}},$
$C_{22}=\frac{0.1}{x_{1}}+\frac{0.1}{x_{2}}+\frac{0.2}{x_{3}}+\frac{0}{x_{4}}$
and
$C_{23}=\frac{0.1}{x_{1}}+\frac{0}{x_{2}}+\frac{0.4}{x_{3}}+\frac{0.5}{x_{4}}.$
According to Definition 3.1, we have that
$C_{x_{1}}=\frac{0.1}{x_{1}}
+\frac{0}{x_{2}}+\frac{0.2}{x_{3}}+\frac{0}{x_{4}},$
$C_{x_{2}}=\frac{0.1}{x_{1}}
+\frac{0.1}{x_{2}}+\frac{0.2}{x_{3}}+\frac{0}{x_{4}},$
$C_{x_{3}}=\frac{0.1}{x_{1}}
+\frac{0}{x_{2}}+\frac{0.2}{x_{3}}+\frac{0}{x_{4}}$ and
$C_{x_{4}}=\frac{0.1}{x_{1}}
+\frac{0}{x_{2}}+\frac{0.4}{x_{3}}+\frac{0.5}{x_{4}}.$ Taking
$X=\frac{0.2}{x_{1}}
+\frac{0.5}{x_{2}}+\frac{0.6}{x_{3}}+\frac{0}{x_{4}}$, according to
Definition 3.8, it follows that
$\underline{X}_{\mathscr{C}}=\frac{0.2}{x_{1}}
+\frac{0.4}{x_{2}}+\frac{0.5}{x_{3}}+\frac{0}{x_{4}}$ and
$\underline{X}_{Cov(\mathscr{C})}=\frac{0.1}{x_{1}}
+\frac{0.1}{x_{2}}+\frac{0.2}{x_{3}}+\frac{0}{x_{4}}$. Consequently,
$\underline{X}_{\mathscr{C}} \nsubseteq
\underline{X}_{Cov(\mathscr{C})}.$ Similarly, we obtain that
$\overline{X}_{\mathscr{C}} \neq \overline{X}_{Cov(\mathscr{C})}.$
\end{upshape}\end{example}

It is well known that the upper approximation can be represented
with neighborhoods in the classical covering approximation space.
But we do not have the same result in the fuzzy covering
approximation space.

\begin{theorem}
Let $(U, \mathscr{C})$ be a fuzzy covering approximation space. Then
$\bigcup\{C_{x}|X(x)>0\}\subseteq \overline{X}_{\mathscr{C}}$ holds
for $X\subseteq U$.
\end{theorem}

\noindent\textbf{Proof.} Taking any $X\subseteq U$, according to
Definition 3.8, we see that
$\overline{X}_{\mathscr{C}}=(\bigcup\{C_{x}|X(x)>0\text{ and }
\underline{X}_{\mathscr{C}}(x)=0\})\cup \underline{X}_{\mathscr{C}}$
and $\bigcup\{C_{x}|X(x)>0\}=(\bigcup\{C_{x}|X(x)>0 \text{ and }
\underline{X}_{\mathscr{C}}(x)=0\})\cup(\bigcup\{C_{x}|\underline{X}_
{\mathscr{C}}(x)>0 \})$. It follows that there exist $C\in
\mathscr{C}$ and $C\subseteq X$ for any $x$ satisfying
$\underline{X}(x)>0$. Consequently, $C_{x}\subseteq C\subseteq X$.
Hence, $\bigcup\{C_{x}|\underline{X}_{\mathscr{C}}(x)>0 \}\subseteq
X$. Therefore, $\bigcup\{C_{x}|X(x)>0\}\subseteq
\overline{X}_{\mathscr{C}}$ holds for $X\subseteq U$.       $\Box$

We see that
$\overline{X}_{\mathscr{C}}\subseteq\bigcup\{C_{x}|X(x)>0\}$ does
not necessarily hold for any $X\subseteq U$. So the upper
approximation can not be represented with neighborhoods only in the
fuzzy covering approximation space. To show this point, we give an
example below.

\begin{example}
\begin{upshape}
Let $U=\{x_{1}, x_{2}, x_{3}, x_{4}\}$, and $\mathscr{C}=\{C_{17},
C_{18}, C_{19}, $ $C_{20}\}$.  For
$X=\frac{0.2}{x_{1}}+\frac{0.2}{x_{2}}+\frac{0.2}{x_{3}}+\frac{0.2}{x_{4}}$,
it follows that $\underline{X}_{\mathscr{C}}=\frac{0.2}{x_{1}}
+\frac{0.2}{x_{2}}+\frac{0.2}{x_{3}}+\frac{0.2}{x_{4}}=\overline{X}_{\mathscr{C}}$.
Furthermore, $\bigcup\{C_{x}|X(x)>0\}=\frac{0.1}{x_{1}}
+\frac{0.1}{x_{2}}+\frac{0.1}{x_{3}}+\frac{0.1}{x_{4}}$. Obviously,
$\overline{X}_{\mathscr{C}}\nsubseteq \bigcup\{C_{x}|X(x)>0\}$.
\end{upshape}
\end{example}

We now investigate the relationship between the lower and upper
approximation operators.

\begin{theorem}
Let $U$ be a non-empty universe of discourse, and $\mathscr{C}_{1},
\mathscr{C}_{2}\in C(U)$. If
$\underline{X}_{\mathscr{C}_{1}}=\underline{X}_{\mathscr{C}_{2}}$
for any $X\subseteq U$, then
$\overline{X}_{\mathscr{C}_{1}}=\overline{X}_{\mathscr{C}_{2}}$.
\end{theorem}

\noindent\textbf{Proof.} By Definition 3.1, we have that
$C_{1x}=\bigcap\{C_{i}|C_{i}(x)>0, C_{i}\in \mathscr{C}_{1}, i\in
I\}$ and $C_{2x}=\bigcap\{C_{j}|C_{j}(x)>0, C_{j}\in
\mathscr{C}_{2}, j\in J\}$.  For any $C_{i},$ where $i\in I$,
$C_{i}=\underline{C}_{i\mathscr{C}_{1}}=\underline{C}_{i\mathscr{C}_{2}}$.
So there exists at least $C_{j}\in \mathscr{C}_{2}$ such that
$C_{j}\subseteq C_{i}$ and $C_{j}(x)>0$. Hence, $C_{2x}\subseteq
C_{1x}$. Similarly, we obtain that $C_{1x}\subseteq C_{2x}$.
Therefore,
$\overline{X}_{\mathscr{C}_{1}}=\overline{X}_{\mathscr{C}_{2}}$.
$\Box$

From Theorem 3.18, we see that the lower and upper approximation
operations are not independent in the fuzzy covering approximation
space. Concretely, the lower approximation operation dominates the
upper one.

\begin{theorem}
Let $U$ be a non-empty universe of discourse, and $\mathscr{C}_{1},
\mathscr{C}_{2}\in C(U)$. Then
$\underline{C}_{\mathscr{C}_{1}}=\underline{C}_{\mathscr{C}_{2}}$
holds for any $C\in \mathscr{C}_{1}\cup \mathscr{C}_{2}$ if and only
if $\underline{X}_{\mathscr{C}_{1}}=\underline{X}_{\mathscr{C}_{2}}$
for any $X\subseteq U$.
\end{theorem}
\noindent\textbf{Proof.} Taking any $X\subseteq U$, by Definition
3.8, it follows that
$\underline{X}_{\mathscr{C}_{1}}=\bigcup\{C_{i}|C_{i}\subseteq X,
C_{i}\in \mathscr{C}_{1}, i\in I\}$. For any $C_{i}\subseteq
\underline{X}_{\mathscr{C}_{1}}$, we have that
$C_{i}=\underline{C}_{i\mathscr{C}_{1}}=\underline{C}_{i\mathscr{C}_{2}}
=\bigcup\{C_{ij}|C_{ij}\in \mathscr{C}_{2},C_{ij}\subseteq X, i\in
I, j\in J\}$. It implies that
$\underline{X}_{\mathscr{C}_{1}}\subseteq
\underline{X}_{\mathscr{C}_{2}}$. Analogously, it follows that
$\underline{X}_{\mathscr{C}_{2}}\subseteq
\underline{X}_{\mathscr{C}_{1}}$. Thereby,
$\underline{X}_{\mathscr{C}_{1}}=\underline{X}_{\mathscr{C}_{2}}$
for any $X\subseteq U$.

The converse is obvious by Definition 3.8. $\Box$

This result indicates that each elementary set in a fuzzy covering
is definable in the other fuzzy covering if and only if two fuzzy
coverings of a universe give the same lower approximations.

\subsection{The fuzzy subcovering and its properties}

It is well-known that the classical upper approximation based on
neighborhoods can be defined equivalently by using a family of
subcoverings. In this subsection, we propose the notion of fuzzy
subcoverings and investigate the relationship between the upper
approximation based on neighborhoods and subcoverings in the fuzzy
covering approximation space.

\begin{definition}
Let $(U, \mathscr{C})$ be a fuzzy covering approximation space,
$X\subseteq U$, and $\mathscr{C}'\subseteq \mathscr{C}$. If
$X\subseteq \bigcup\{C|C\in \mathscr{C}'\}$ , then $\mathscr{C}'$ is
called a fuzzy subcovering of $X$.
\end{definition}

In other words, the fuzzy subcovering of $X$ is a subset of
$\mathscr{C}$ which covers $X$. Obviously, $\mathscr{C}$ is the
maximum fuzzy subcovering for $X\subseteq U$ if $X\subseteq \bigcup
\mathscr{C}$. In this work, we denote the set of all the fuzzy
subcoverings of $X$ as $FC(X)$.

\begin{theorem}
Let $(U, \mathscr{C})$ be a fuzzy covering approximation space. Then
$\overline{X}_{\mathscr{C}}\subseteq \bigcap\{\bigcup\{C|C\in
\mathscr{C}'\}| \mathscr{C}'\in FC(X)\}$ holds for any $X\subseteq
U$.
\end{theorem}

\noindent\textbf{Proof.} Taking any $X\subseteq U$, by Definition
3.8, it follows that
$\overline{X}_{\mathscr{C}}=(\bigcup\{C_{x}|X(x)>0\text{ and }
\underline{X}_{\mathscr{C}}(x)=0\})\cup
\underline{X}_{\mathscr{C}}$. By Proposition 3.9, it implies that
$\underline{X}\subseteq X\subseteq \bigcup\{C|C\in \mathscr{C}'\in
FC(X)\}$. Evidently, $C_{x}\subseteq \bigcup\{C|C\in \mathscr{C}'\in
FC(X)\}$ for any $x\in U$ satisfying $X(x)>0$. Thereby,
$\overline{X}_{\mathscr{C}}\subseteq \bigcap\{\bigcup\{C|C\in
\mathscr{C}'\}| \mathscr{C}'\in FC(X)\}$ holds for any $X\subseteq
U$.       $\Box$

However, $\bigcap\{\bigcup\{C|C\in \mathscr{C}'\}| \mathscr{C}'\in
FC(X)\}\subseteq \overline{X}_{\mathscr{C}}$ does not hold
generally. That is, the upper approximation may not be represented
with a family of fuzzy subcoverings of $X$ as the classical covering
approximation space, which is shown by the following example.

\begin{example}\begin{upshape}
Let $U=\{x_{1}, x_{2}, x_{3}, x_{4}\}$, $\mathscr{C}=\{C_{24},
C_{25}, C_{26}\}$, where
$C_{24}=\frac{0.2}{x_{1}}+\frac{0.1}{x_{2}}+\frac{0.2}{x_{3}}+\frac{0.1}{x_{4}},$
$C_{25}=\frac{0.1}{x_{1}}+\frac{0.2}{x_{2}}+\frac{0.1}{x_{3}}+\frac{0.2}{x_{4}}$
and
$C_{26}=\frac{0.1}{x_{1}}+\frac{0.1}{x_{2}}+\frac{0.2}{x_{3}}+\frac{0.1}{x_{4}}$.
According to Definition 3.20, we obtain all fuzzy subcoverings of
$X=
\frac{0.1}{x_{1}}+\frac{0}{x_{2}}+\frac{0.2}{x_{3}}+\frac{0}{x_{4}}$
as $FC(X)=\{\{C_{24}\},\{C_{26} \},\{C_{24}, C_{25}\},\{C_{24},
C_{26}\},$ $\{C_{25}, C_{26}\},$ $\{C_{24}, C_{25}, C_{26}\}\}$. It
follows that $\bigcap\{\bigcup\{C|C\in \mathscr{C}'\}|
\mathscr{C}'\in
FC(X)\}=\frac{0.1}{x_{1}}+\frac{0.1}{x_{2}}+\frac{0.2}{x_{3}}+\frac{0.1}{x_{4}}$,
but $\overline{X}_{\mathscr{C}}=\frac{0.1}{x_{1}}
+\frac{0.1}{x_{2}}+\frac{0.1}{x_{3}}+\frac{0.1}{x_{4}}$. Therefore,
$\bigcap\{\bigcup\{C|C\in \mathscr{C}'\}| \mathscr{C}'\in
FC(X)\}\nsubseteq\overline{X}_{\mathscr{C}}$.
\end{upshape}\end{example}

According to Theorems 3.16 and 3.21, we have that
$\bigcup\{C_{x}|X(x)>0\}\subseteq
\overline{X}_{\mathscr{C}}\subseteq \bigcap\{\bigcup\{C|C\in
\mathscr{C}'\}| \mathscr{C}'\in FC(X)\}$ for any $X\subseteq U$.

Sometimes, the fuzzy covering $\mathscr{C}$ of $U$ is a trivial
subcovering of $X\subseteq U$. Specially, we do not take
$\mathscr{C}$ into account in the following situation.

\begin{proposition}
Let $(U, \mathscr{C})$ be a fuzzy covering approximation space, and
$X\subseteq U$. If $|FC(X)|\geq 2$, where $|FC(X)|$ stands for the
cardinality of $FC(X)$, then $\bigcap\{\bigcup\{C|C\in
\mathscr{C}'\}| \mathscr{C}'\in FC(X)\}= \bigcap\{\bigcup\{C|C\in
\mathscr{C}'\}| \mathscr{C}'\in FC(X)-\{\mathscr{C}\}\}$.
\end{proposition}

\noindent\textbf{Proof.} Straightforward.       $\Box$

\subsection{ The irreducible and reducible elements of a fuzzy covering}

In this subsection, we provide the concepts of reducible and
irreducible elements to formally investigate the relationship among
elementary elements of a fuzzy covering. Although several theorems
in this subsection are special cases of \cite{Zhang1}, they don't
give their proofs. To better understand the following results, we
prove them concretely in the following.

\begin{definition}\cite{Zhang1}
Let $(U, \mathscr{C})$ be a fuzzy covering approximation space, and
$C\in \mathscr{C}$.  If $C$ can not be written as a union of some
sets in $\mathscr{C}-\{C\}$, then $C$ is called an irreducible
element. Otherwise, $C$ is called a reducible element.
\end{definition}

It is obvious that the concept of the irreducible element in a fuzzy
covering approximation space is an extension of the notion of the
irreducible element in a covering approximation space, and the
irreducible element can be used for the definition of reducts of
fuzzy coverings.

\begin{proposition}\cite{Zhang1}
Let $(U, \mathscr{C})$ be a fuzzy covering approximation space, and
$C$ a reducible element of $\mathscr{C}$. Then $\mathscr{C}-\{C\}$
is still a fuzzy covering of $U$.
\end{proposition}

In other words, a fuzzy covering  of a universe deleting all
reducible elements is a fuzzy covering, and the rest elements are
irreducible.

\begin{definition}\cite{Zhang1}
Let $(U, \mathscr{C})$ be a fuzzy covering approximation space. If
every element of $\mathscr{C}$ is an irreducible element, then
$\mathscr{C}$ is irreducible. Otherwise, $\mathscr{C}$ is reducible.
\end{definition}

Next, we discuss the properties of reducible elements of a fuzzy
covering.

\begin{theorem}\cite{Zhang1}
Let $(U, \mathscr{C})$ be a fuzzy covering approximation space, $C$
a reducible element of $\mathscr{C}$, and $C_{0}\in
\mathscr{C}-\{C\}$. Then $C_{0}$ is a reducible element of
$\mathscr{C}$ if and only if it is a reducible element of
$\mathscr{C}-\{C\}$.
\end{theorem}

\noindent\textbf{Proof.} We assume that $C_{0}$ is a reducible
element of $\mathscr{C}$. It follows that we can express $C_{0}$ as
a union of subset of $\mathscr{C}-\{C_{0}\}$, denoted as $C_{1},
C_{2},...,C_{N}$. If there exists no set which is equal to $C$ in
$\{C_{1}, C_{2},...,C_{N}\}$, then $C_{0}$ is a reducible element of
$\mathscr{C}-\{C\}$. If there is a set which is equal to $C$ in
$\{C_{1}, C_{2},...,C_{N}\}$, taking $C_{1}=C$, then $C_{1}$ is the
union of some sets $\{D_{1}, D_{2},...,D_{M}\}$ in
$\mathscr{C}-\{C\}$. Consequently, we obtain that $C_{0}=D_{1}\cup
D_{2}\cup...D_{M}\cup C_{2}\cup...\cup C_{N}$. Clearly, $D_{1},
D_{2},...,D_{M}, C_{2},...,C_{N}$ are not equal to either $C_{0}$ or
$C$. So $C$ is a reducible element of $\mathscr{C}-\{C\}$.

Since $C_{0}$ is a reducible element of $\mathscr{C}-\{C\}$, it can
be expressed as a union of some sets in $\mathscr{C}-\{C, C_{0}\}$.
We can express it as a union of some sets in
$\mathscr{C}-\{C_{0}\}$. Therefore, $C_{0}$ is a reducible element
of $\mathscr{C}$. $\Box$

Next, we investigate the relationship between the approximation
operations and the reducible elements in the fuzzy covering
approximation space.

\begin{theorem}\cite{Zhang1}
Let $(U, \mathscr{C})$ be a fuzzy covering approximation space, and
$C$ a reducible element of $\mathscr{C}$. Then
$\underline{X}_{\mathscr{C}}=\underline{X}_{\mathscr{C}-\{C\}}$
holds for any $X\subseteq U$.
\end{theorem}

\noindent\textbf{Proof.} Taking any $X\subseteq U$, by Definition
3.24, it follows that $\underline{X}_{\mathscr{C}-\{C\}}\subseteq
\underline{X}_{\mathscr{C}}\subseteq X$. Moreover, there exist
$C_{1}, C_{2},...,C_{N}$ such that
$\underline{X}_{\mathscr{C}}=C_{1}\cup C_{2}\cup...\cup C_{N}$. If
none of $C_{1}, C_{2},...,C_{N}$ is equal to $C$, then they belong
to $\mathscr{C}-\{C\}$. Consequently, $C_{1}, C_{2},...,C_{N}$ are
all the subsets of $\underline{X}_{\mathscr{C}-\{C\}}$. If there is
a set which is equal to $C$, then we take $C=C_{1}$. Since $C$ is a
reducible element of $\mathscr{C}$, $C$ can be expressed as some
sets in $\mathscr{C}-\{C\}$ such that $C=D_{1}\cup D_{2}\cup...\cup
D_{M}$. Hence, $\underline{X}_{\mathscr{C}}=D_{1}\cup
D_{2}\cup...\cup D_{M}\cup C_{2}\cup...\cup C_{N}$. It implies that
$\underline{X}_{\mathscr{C}}\subseteq
\underline{X}_{\mathscr{C}-\{C\}}$. Therefore,
$\underline{X}_{\mathscr{C}}=\underline{X}_{\mathscr{C}-\{C\}}$
holds for any $X\subseteq U$.       $\Box$

In other words, the lower approximation of any $X\subseteq U_{1}$ in
$\mathscr{C}$ is the same as that in $\mathscr{C}-\{C\}$ if $C$ is
reducible.

\begin{corollary}\cite{Zhang1}
Let $(U, \mathscr{C})$ be a fuzzy covering approximation space, and
$C$ a reducible element of $\mathscr{C}$. Then
$\overline{X}_{\mathscr{C}}=\overline{X}_{\mathscr{C}-\{C\}}$ holds
for any $X\subseteq U$.
\end{corollary}

\noindent\textbf{Proof.} Straightforward from Theorems 3.27 and
3.28. $\Box$

In this sequel, we use $RED(\mathscr{C})$ to represent the set of
all irreducible elements of a fuzzy covering $\mathscr{C}$. It is
easy to see that $RED(\mathscr{C})=RED(RED(\mathscr{C}))$. Next, we
study the relationship between $RED(\mathscr{C})$ and the lower and
upper approximation operations.

\begin{corollary}
Let $(U, \mathscr{C})$ be a fuzzy covering approximation space. Then
$\underline{X}_{\mathscr{C}}=\underline{X}_{RED(\mathscr{C})}$ holds
for any $X\subseteq U$.
\end{corollary}

\noindent\textbf{Proof.} Straightforward from Theorem 3.28. $\Box$

\begin{corollary}
Let $(U, \mathscr{C})$ be a fuzzy covering approximation space. Then
$\overline{X}_{\mathscr{C}}=\overline{X}_{RED(\mathscr{C})}$ holds
for any $X\subseteq U$.
\end{corollary}

\noindent\textbf{Proof.} Straightforward from Corollary 3.29. $\Box$

Based on Theorem 3.28, Corollaries 3.29, 3.30 and 3.31, we obtain
the following theorem.

\begin{theorem}
Let $U$ be a universe, $\mathscr{C}_{1}$ and $\mathscr{C}_{2}$ two
irreducible fuzzy coverings of $U$. If
$\underline{X}_{\mathscr{C}_{1}}=\underline{X}_{\mathscr{C}_{2}}$
for any $X\subseteq U$. then the two fuzzy coverings are the same.
\end{theorem}

\noindent\textbf{Proof.}  Taking any $C\in \mathscr{C}_{1}$, by
Definition 3.8, it follows that
$\underline{C}_{\mathscr{C}_{1}}=C=\underline{C}_{\mathscr{C}_{2}}$.
Consequently, $C$ is the union of some sets of $\mathscr{C}_{2}$
such that $C=C_{1}\cup C_{2}\cup...\cup C_{N}$. Similarly, there
exist $D_{i1}, D_{i2},...,D_{iM(i)}\in \mathscr{C}_{1}$ such that
$C_{i}=D_{i1}\cup D_{i2}\cup...\cup D_{iM(i)}.$ Hence, $C=D_{11}\cup
D_{12}\cup...\cup D_{N1}\cup D_{N2}\cup...\cup D_{NM(N)}$. Since $C$
is irreducible, $C=D_{ij}$ for all $i, j$. It implies that $C$ is an
element of $\mathscr{C}_{2}$. On the other hand, any element of
$\mathscr{C}_{2}$ is an element of $\mathscr{C}_{1}$. Therefore, the
two fuzzy coverings $\mathscr{C}_{1}$ and $\mathscr{C}_{2}$ are the
same.        $\Box$

\begin{corollary}
Let $U$ be a universe, $\mathscr{C}_{1}$ and $\mathscr{C}_{2}$ two
irreducible fuzzy coverings of $U$. If
$\overline{X}_{\mathscr{C}_{1}}=\overline{X}_{\mathscr{C}_{2}}$ for
 any $X\subseteq U$, then the two fuzzy coverings are the same.
\end{corollary}

\noindent\textbf{Proof.} The proof is similar to that in Theorem
3.32. $\Box$

\begin{theorem}
Let $U$ be a non-empty universe of discourse, and $\mathscr{C}_{1},
\mathscr{C}_{2}\in C(U)$. Then
$\underline{X}_{\mathscr{C}_{1}}=\underline{X}_{\mathscr{C}_{2}}$
holds for any $X\subseteq U$ if and only if
$RED(\mathscr{C}_{1})=RED(\mathscr{C}_{2})$.
\end{theorem}

\noindent\textbf{Proof.} Since
$\underline{X}_{\mathscr{C}_{1}}=\underline{X}_{\mathscr{C}_{2}}$
for any $X\subseteq U$,
$\underline{C}_{\mathscr{C}_{1}}=\underline{C}_{\mathscr{C}_{2}}$
for any $C\in \mathscr{C}_{1}\cup \mathscr{C}_{2}$. Taking any $C\in
RED(\mathscr{C}_{1})$, it follows that $C=\bigcup\{C_{i}|C_{i}\in
RED(\mathscr{C}_{2}), i\in I\}=\bigcup\{\bigcup\{C_{ij}|C_{ij}\in
\mathscr{C}_{1},i\in I \}|j\in J\}$. It implies that $C\in
\mathscr{C}_{2}$. Hence, $RED(\mathscr{C}_{1})\subseteq
RED(\mathscr{C}_{2})$. Similarly, we obtain that
$RED(\mathscr{C}_{2})\subseteq RED(\mathscr{C}_{1})$. Therefore,
$RED(\mathscr{C}_{1})= RED(\mathscr{C}_{2})$.

The converse is obvious by Definitions 3.8 and 3.24.       $\Box$

It can be seen from Theorem 3.34 that two fuzzy coverings of a
universe generate the same lower approximation if and only if there
exist the same irreducible elements in these fuzzy coverings.

To illustrate Theorem 3.34, we supply the following example.

\begin{example}\begin{upshape}
Let $U=\{x_{1}, x_{2}, x_{3}, x_{4}\}$, $\mathscr{C}_{1}=\{C_{17},
C_{18},$ $ C_{19}, C_{20}, C_{27}, C_{28}\}$,
$\mathscr{C}_{2}=\{C_{17}, C_{18}, C_{19}, C_{20}, C_{29},$ $
C_{30}\}$, where
$C_{27}=\frac{0.2}{x_{1}}+\frac{0.2}{x_{2}}+\frac{0.1}{x_{3}}+\frac{0.1}{x_{4}},$
$C_{28}=\frac{0.2}{x_{1}}+\frac{0.1}{x_{2}}+\frac{0.2}{x_{3}}+\frac{0.1}{x_{4}},$
$C_{29}=\frac{0.1}{x_{1}}+\frac{0.1}{x_{2}}+\frac{0.2}{x_{3}}+\frac{0.2}{x_{4}}$
and
$C_{30}=\frac{0.1}{x_{1}}+\frac{0.2}{x_{2}}+\frac{0.1}{x_{3}}+\frac{0.2}{x_{4}}.$
Obviously, we obtain that $RED(\mathscr{C}_{1})=
RED(\mathscr{C}_{2})=\{C_{17}, C_{18}, C_{19}, C_{20}\}$.
\end{upshape}\end{example}

\begin{corollary}
Let $U$ be a non-empty universe of discourse, and $\mathscr{C}_{1},
\mathscr{C}_{2}\in C(U)$. If
$\overline{X}_{\mathscr{C}_{1}}=\overline{X}_{\mathscr{C}_{2}}$ for
any $X\subseteq U$ if and only if
$RED(\mathscr{C}_{1})=RED(\mathscr{C}_{2})$.
\end{corollary}

From Corollary 3.36, we see that they have the same irreducible
elements if and only if two fuzzy coverings of a universe generate
the same upper approximation.

\begin{corollary}
Let $(U, \mathscr{C}_{1})$ be a fuzzy covering approximation space,
$\mathscr{C}_{2}=\{\bigcup_{C\in \mathscr{C}'}C|\emptyset\neq
\mathscr{C}'\subseteq \mathscr{C}_{1}\}$, and $X\subseteq U$. Then
$\underline{X}_{\mathscr{C}_{1}}=\underline{X}_{\mathscr{C}_{2}}$
and $\overline{X}_{\mathscr{C}_{1}}=\overline{X}_{\mathscr{C}_{2}}$.
\end{corollary}

\noindent\textbf{Proof.} By Definition 3.24, we observe that
$RED(\mathscr{C}_{1})=RED(\mathscr{C}_{2})$. Therefore,
$\underline{X}_{\mathscr{C}_{1}}=\underline{X}_{\mathscr{C}_{2}}$
and $\overline{X}_{\mathscr{C}_{1}}=\overline{X}_{\mathscr{C}_{2}}$.
$\Box$

We also investigate the relationship between the reducible elements
and the neighborhood operator in the fuzzy covering approximation
space.

\begin{theorem}
Let $(U, \mathscr{C})$ be a fuzzy covering approximation space, and
$C$ a reducible element of $\mathscr{C}$.  Then $C_{x}$ in
$\mathscr{C}-\{C\}$ is the same as that in $\mathscr{C}$ for any
$x\in U$.
\end{theorem}

\noindent\textbf{Proof.} By Definitions 3.1 and 3.24, we have that
$C_{x}=\bigcap\{C_{i}|C_{i}(x)>0, C_{i}\in
\mathscr{C}\}=\bigcap\{C_{i}|C_{i}(x)>0, C_{i}\in
\mathscr{C}-\{C\}\}$ for any $x\in U$. Therefore, $C_{x}$ in
$\mathscr{C}-\{C\}$ is the same as that in $\mathscr{C}$ for any
$x\in U$.       $\Box$

That is to say, if we delete some reducible elements in the fuzzy
covering, then it will not change the neighborhood $C_{x}$ for any
$x\in U$.

\begin{corollary}
Let $(U, \mathscr{C})$ be a fuzzy covering approximation space. Then
$C_{x}$ in $RED(\mathscr{C})$ is the same as that in $\mathscr{C}$
for any $x\in U$.
\end{corollary}

\noindent\textbf{Proof.} Straightforward from Theorem 3.38. $\Box$

Corollary 3.39 indicates that $RED(\mathscr{C})$ and $\mathscr{C}$
generate the same neighborhood $C_{x}$ for any $x\in U$ in the fuzzy
covering approximation space.

\begin{corollary}
Let $U$ be a non-empty universe of discourse, and $\mathscr{C}_{1},
\mathscr{C}_{2}\in C(U)$. If
$RED(\mathscr{C}_{1})=RED(\mathscr{C}_{2})$, then $C_{x}$ in
$\mathscr{C}_{1}$ is the same as that in $\mathscr{C}_{2}$ for any
$x\in U$.
\end{corollary}

\noindent\textbf{Proof.} Straightforward from Corollary 3.39. $\Box$

By Corollary 3.40, if there exist the same irreducible elements in
two fuzzy coverings $\mathscr{C}_{1}$ and $\mathscr{C}_{2}$ of $U$,
then they generate the same neighborhood $C_{x}$ for any $x\in U$.

\begin{theorem}
Let $U$ be a non-empty universe of discourse, and $\mathscr{C}_{1},
\mathscr{C}_{2}\in C(U)$. If
$\overline{C}_{\mathscr{C}_{1}}=\overline{C}_{\mathscr{C}_{2}}$ for
any $C\in \mathscr{C}_{1}\cup \mathscr{C}_{2}$, then
$\bigcup\{C_{1x}|X(x)>0\} =\bigcup\{C_{2x}|X(x)>0\}$ for any
$X\subseteq U$.
\end{theorem}

\noindent\textbf{Proof.} By Definition 3.1, we have that
$C_{1x}=\bigcap\{C_{i}|C_{i}(x)>0, C_{i}\in \mathscr{C}_{1}, i\in
I\}$ and $C_{2x}=\bigcap\{C_{j}|C_{j}(x)>0, C_{j}\in
\mathscr{C}_{2}, j\in J\}$ for any $x\in U$. Assume that there
exists $x \in U$ such that $X(x)>0$ and $C_{1x} \neq C_{2x}$.
Without loss of generality, there is $y\in U$ such that
$(C_{1x})(y)>0$ and $(C_{2x})(y)=0$. Obviously, $y\neq x$. Hence,
there exist $C_{j}(y)=0$ and $C_{j}(x)>0$. But
$C_{j}=\overline{C}_{j\mathscr{C}_{2}}=\overline{C}_{j\mathscr{C}_{1}}\supseteq
\bigcup\{C_{1z}|C_{j}(z)>0\}\supseteq C_{1x}$. It implies that
$C_{j}(y)>0$, which is a contradiction. Consequently, $C_{1x} =
C_{2x}$ for any $x\in U$. Therefore, $\bigcup\{C_{1x}|X(x)>0\}
=\bigcup\{C_{2x}|X(x)>0\}$ for any $X\subseteq U$.       $\Box$

Theorem 3.41 shows that two fuzzy coverings of a universe generate
the same neighborhood $C_{x}$ for any $x\in U$ if each elementary
element has the same lower approximation in two fuzzy coverings.

\subsection{The non-intersectional and intersectional
elements of fuzzy coverings, the union and intersection operations
on fuzzy coverings}

For any universal set $U$, we denote $CC(U)$ as the set of all
coverings of $U$. It is well-known that the number of possible
coverings for a set $U$ of $n$ elements is
$$|CC(U)|=\frac{1}{2}\sum^{n}_{k=0}(\frac{n}{k})2^{2^{n-k}},$$
the first few of which are 1, 5, 109, 32297, 2147321017. Since
$C(U)$ contains a larger number of fuzzy coverings than $CC(U)$ in
practice, it is of interest to investigate the relationship between
fuzzy coverings. In this subsection, we introduce several operations
on fuzzy coverings and study their basic properties for facilitating
the computation of fuzzy coverings.

\begin{definition}
Let $(U, \mathscr{C})$ be a fuzzy covering approximation space, and
$C\in \mathscr{C}$. If $C$ can not be written as an intersection of
some sets in $\mathscr{C}-\{C\}$, then $C$ is called a
non-intersectional element. Otherwise, $C$ is called an
intersectional element.
\end{definition}

For simplicity, we use $IS(\mathscr{C})$ to represent the set of all
non-intersectional elements of $\mathscr{C}$. It is easy to see that
$IS(\mathscr{C})=IS(IS(\mathscr{C}))$. Notice that the function $IS:
C(U)\longrightarrow C(U)$ that maps $\mathscr{C}$ to
$IS(\mathscr{C})$ is well-defined. Hence, we may view $IS$ as a
unary operator on $C(U)$.

We employ an example to illustrate the non-intersectional and
intersectional elements in the following.

\begin{example}\begin{upshape}
Let $U=\{x_{1}, x_{2}, x_{3}, x_{4}\}$, $\mathscr{C}=\{C_{17},
C_{18}, C_{19}, C_{20}, C_{31}\}$, where
$C_{31}=\frac{0.1}{x_{1}}+\frac{0.1}{x_{2}}+\frac{0.1}{x_{3}}+\frac{0.1}{x_{4}}$.
By Definition 3.42, we have that $IS(\mathscr{C})=\{C_{17}, C_{18},
C_{19}, C_{20}\}$. \end{upshape}\end{example}

\begin{proposition}
Let $(U, \mathscr{C})$ be a fuzzy covering approximation space. Then
$\bigcup\{C_{x}|X(x)>0\}=\bigcup\{C_{IS(\mathscr{C})x}|$ $X(x)>0\}$
and $\underline{X}_{IS(\mathscr{C})}\subseteq
\underline{X}_{\mathscr{C}}$ for any $X\subseteq U$.
\end{proposition}

\noindent\textbf{Proof.} By Definition 3.42, it follows that
$C_{x}=C_{IS(\mathscr{C})x}$ for any $x\in U$. Consequently,
$\bigcup\{C_{x}|X(x)>0\}=\bigcup\{C_{IS(\mathscr{C})x}|X(x)>0\}$ for
any $X\subseteq U$. Furthermore, since $IS(\mathscr{C})\subseteq
\mathscr{C}$, we have that $\underline{X}_{IS(\mathscr{C})}\subseteq
\underline{X}_{\mathscr{C}}$ for any $X\subseteq U$.       $\Box$

We observe that the neighborhood $C_{x}$ generated in the fuzzy
covering $\mathscr{C}$ is the same as that generated in all
non-intersectional elements of $\mathscr{C}$. On the other hand, $
\overline{X}_{\mathscr{C}}\subseteq \overline{X}_{IS(\mathscr{C})}$
does not necessarily hold for any $X\subseteq U$. An example is
given to illustrate this point.

\begin{example}\begin{upshape}
Let $U=\{x_{1}, x_{2}, x_{3}, x_{4}\}$, $\mathscr{C}=\{C_{32},
C_{33},$ $ C_{34}, C_{35}, C_{36}, C_{37}, C_{38}, C_{39}\}$, where
$C_{32}=\frac{0.1}{x_{1}}+\frac{0}{x_{2}}+\frac{0}{x_{3}}+\frac{0}{x_{4}},$
$C_{33}=\frac{0}{x_{1}}+\frac{0.1}{x_{2}}+\frac{0}{x_{3}}+\frac{0}{x_{4}},$
$C_{34}=\frac{0.1}{x_{1}}+\frac{0.2}{x_{2}}+\frac{0}{x_{3}}+\frac{0}{x_{4}},$
$C_{35}=\frac{0.1}{x_{1}}+\frac{0}{x_{2}}+\frac{0.1}{x_{3}}+\frac{0}{x_{4}},$
$C_{36}=\frac{0.4}{x_{1}}+\frac{0.2}{x_{2}}+\frac{0.1}{x_{3}}+\frac{0}{x_{4}},$
$C_{37}=\frac{0.1}{x_{1}}+\frac{0.2}{x_{2}}+\frac{0}{x_{3}}+\frac{0.1}{x_{4}},$
$C_{38}=\frac{0.1}{x_{1}}+\frac{0}{x_{2}}+\frac{0.1}{x_{3}}+\frac{0.5}{x_{4}}$
and
$C_{39}=\frac{0}{x_{1}}+\frac{0.1}{x_{2}}+\frac{0.4}{x_{3}}+\frac{0.4}{x_{4}}.$
Evidently, $IS(\mathscr{C})=\{C_{36}, C_{37}, C_{38}, C_{39}\}$.
Taking
$X=\frac{0.4}{x_{1}}+\frac{0.2}{x_{2}}+\frac{0}{x_{3}}+\frac{0}{x_{4}},$
according to Definitions 3.8 and 3.42, we obtain that
$\overline{X}_{\mathscr{C}}=\frac{0.1}{x_{1}}
+\frac{0.2}{x_{2}}+\frac{0}{x_{3}}+\frac{0}{x_{4}}$ and
$\overline{X}_{IS(\mathscr{C})}=\frac{0.1}{x_{1}}
+\frac{0.1}{x_{2}}+\frac{0}{x_{3}}+\frac{0}{x_{4}}.$ Thereby,
$\overline{X}_{\mathscr{C}}\nsubseteq
\overline{X}_{IS(\mathscr{C})}$. \end{upshape}\end{example}

Following, we present a theorem for the intersection element of a
fuzzy covering.

\begin{theorem}
Let $(U, \mathscr{C})$ be a fuzzy covering approximation space, $C$
an intersection element of $\mathscr{C}$, and $C_{0}\in
\mathscr{C}-\{C\}$. Then $C_{0}$ is an intersection element of
$\mathscr{C}$ if and only if it is an intersection element of
$\mathscr{C}-\{C\}$.
\end{theorem}

\noindent\textbf{Proof.} The proof is similar to that in Theorem
3.27. $\Box$

Next, we present the notions of the union and intersection
operations on fuzzy coverings, and investigate their basic
properties.

\begin{definition}
Let $U$ be a non-empty universe of discourse, and $\mathscr{C}_{1},
\mathscr{C}_{2}\in C(U)$. If
\begin{equation*} \mathscr{C}_{1}\cup \mathscr{C}_{2}=\{C| C\in \mathscr{C}_{1}
\text{ or } C\in \mathscr{C}_{2}\},
\end{equation*}
then $\mathscr{C}_{1}\cup \mathscr{C}_{2}$ is called the union of
$\mathscr{C}_{1}$ and  $\mathscr{C}_{2}$.
\end{definition}

It is obvious that the union operation is to collect all elementary
elements in each fuzzy covering.

\begin{proposition}
Let $U$ be a non-empty universe of discourse, $\mathscr{C}_{1},
\mathscr{C}_{2}\in C(U)$, and $X\subseteq U$. Then
$\underline{X}_{\mathscr{C}_{i}}
\subseteq\underline{X}_{\mathscr{C}_{1}\bigcup\mathscr{C}_{2}}$,
where $ i=1, 2.$
\end{proposition}

\noindent\textbf{Proof.} According to Definition 3.8, we have that
$\underline{X}_{\mathscr{C}_{i}}=\bigcup\{C\in
\mathscr{C}_{i}|C\subseteq X\}\subseteq (\bigcup\{C\in
\mathscr{C}_{1}|C\subseteq X\})\bigcup (\bigcup\{C\in
\mathscr{C}_{2}|C\subseteq
X\})=\underline{X}_{\mathscr{C}_{1}\bigcup\mathscr{C}_{2}}$.
Thereby, $\underline{X}_{\mathscr{C}_{i}}
\subseteq\underline{X}_{\mathscr{C}_{1}\bigcup\mathscr{C}_{2}}$,
where $ i=1, 2.$        $\Box$

The following example shows that the converse of Proposition 3.48
does not hold generally.

\begin{example}\begin{upshape}
Let $U=\{x_{1}, x_{2}, x_{3}, x_{4}\}$, $\mathscr{C}_{1}=\{C_{21},
C_{22},$ $ C_{23}\}$, and $\mathscr{C}_{2}=\{C_{1x_{1}},
C_{1x_{2}},C_{1x_{3}},C_{1x_{4}}\}$. Taking $X=
\frac{0.2}{x_{1}}+\frac{0.5}{x_{2}}+\frac{0.6}{x_{3}}+\frac{0.1}{x_{4}}$.
By Definition 3.8, we have that $\overline{X}_{\mathscr{C}_{1}}=
\frac{0.2}{x_{1}}+\frac{0.4}{x_{2}}+\frac{0.5}{x_{3}}+\frac{0.5}{x_{4}}$,
$\overline{X}_{\mathscr{C}_{2}}=
\frac{0.1}{x_{1}}+\frac{0.1}{x_{2}}+\frac{0.4}{x_{3}}+\frac{0.5}{x_{4}}$,
$\underline{X}_{\mathscr{C}_{1}\bigcup \mathscr{C}_{2}}=
\frac{0.2}{x_{1}}+\frac{0.4}{x_{2}}+\frac{0.5}{x_{3}}+\frac{0}{x_{4}}$
and $\overline{X}_{\mathscr{C}_{1}\bigcup \mathscr{C}_{2}}=
\frac{0.2}{x_{1}}+\frac{0.4}{x_{2}}+\frac{0.5}{x_{3}}+\frac{0.5}{x_{4}}$.
Obviously, $\overline{X}_{\mathscr{C}_{1}\bigcup
\mathscr{C}_{2}}\nsubseteq \overline{X}_{\mathscr{C}_{2}}$.
\end{upshape}\end{example}

\begin{definition}
Let $U$ be a non-empty universe of discourse, and $\mathscr{C}_{1},
\mathscr{C}_{2}\in C(U)$. If
\begin{equation*} \mathscr{C}_{1}\cap \mathscr{C}_{2}=\{C_{1x}\cap
C_{2x}| C_{ix}\in Cov(\mathscr{C}_{i}), x\in U_{1}, i=1,2 \},
\end{equation*}
then $\mathscr{C}_{1}\cap \mathscr{C}_{2}$ is called the
intersection of $\mathscr{C}_{1}$ and  $\mathscr{C}_{2}$.
\end{definition}

It is obvious that $C_{1x}=\bigcap\{C |C(x)>0, C\in
\mathscr{C}_{1}\}$ and $C_{2x}=\bigcap\{C' |C'(x)>0, C'\in
\mathscr{C}_{2}\}$ for any $x\in U_{1}$. So $\mathscr{C}_{1}\bigcap
\mathscr{C}_{2}$ is a fuzzy covering of $U_{1}$. Furthermore, if we
take the value of membership degree from the set $\{0, 1\}$, then
Definition 3.50 is the same as that in Definition 4.2 in
\cite{Wang2}.

It can be found that $\overline{X}_{\mathscr{C}_{i}}\subseteq
\overline{X}_{\mathscr{C}_{1}\bigcap\mathscr{C}_{2}}$ does not
necessarily hold for any $X\subseteq U$ in the fuzzy covering
approximation space, where $i=1,2$. To illustrate this point, we
give the following example.

\begin{example}
\begin{upshape}
Let $U=\{x_{1}, x_{2}, x_{3}, x_{4}\}$, $\mathscr{C}_{1}=\{C_{21},
C_{22},$ $ C_{23}\}$, and
$\mathscr{C}_{2}=\{\frac{0.2}{x_{1}}+\frac{0.1}{x_{2}}
+\frac{0.4}{x_{3}}+\frac{0.5}{x_{4}}\}$. According to Definition
3.1, we have that $C_{1x_{1}}=\frac{0.1}{x_{1}}+\frac{0}{x_{2}}
+\frac{0.2}{x_{3}}+\frac{0}{x_{4}}$,
$C_{1x_{2}}=\frac{0.1}{x_{1}}+\frac{0.1}{x_{2}}
+\frac{0.2}{x_{3}}+\frac{0}{x_{4}}$,
$C_{1x_{3}}=\frac{0.1}{x_{1}}+\frac{0}{x_{2}}
+\frac{0.2}{x_{3}}+\frac{0}{x_{4}}$ and
$C_{1x_{4}}=\frac{0.1}{x_{1}}+\frac{0}{x_{2}}
+\frac{0.4}{x_{3}}+\frac{0.5}{x_{4}}$. Taking
$X=\frac{0.2}{x_{1}}+\frac{0.4}{x_{2}}
+\frac{0.5}{x_{3}}+\frac{0.5}{x_{4}}$,  it implies that
$\underline{X}_{\mathscr{C}_{1}\bigcap\mathscr{C}_{2}}=\frac{0.1}{x_{1}}
+\frac{0.1}{x_{2}} +\frac{0.4}{x_{3}}+\frac{0.5}{x_{4}}$,
$\underline{X}_{\mathscr{C}_{1}}=\frac{0.2}{x_{1}}+\frac{0.4}{x_{2}}
+\frac{0.4}{x_{3}}+\frac{0.5}{x_{4}}$ and
$\underline{X}_{\mathscr{C}_{2}}=\frac{0.2}{x_{1}}+\frac{0.1}{x_{2}}
+\frac{0.4}{x_{3}}+\frac{0.5}{x_{4}}$. Clearly,
$\overline{X}_{\mathscr{C}_{1}}\nsubseteq
\overline{X}_{\mathscr{C}_{1}\bigcap\mathscr{C}_{2}}$ and
$\overline{X}_{\mathscr{C}_{2}}\nsubseteq
\overline{X}_{\mathscr{C}_{1}\bigcap\mathscr{C}_{2}}$.
\end{upshape}
\end{example}

By Definitions 3.4 and 3.50, we present the following proposition.

\begin{proposition}
Let $U$ be a non-empty universe of discourse, and $\mathscr{C}_{1},
\mathscr{C}_{2}\in C(U)$. Then
$\mathscr{C}_{1}\bigcap\mathscr{C}_{2}=Cov(\mathscr{C}_{1}\bigcup
\mathscr{C}_{2})$.
\end{proposition}

\begin{definition}
Let $U$ be a non-empty universe of discourse, and $\mathscr{C}_{1},
\mathscr{C}_{2}\in C(U)$. If there exists $C^{\ast}\in
\mathscr{C}_{2}$ such that $C\subseteq C^{\ast}$ for any $C\in
\mathscr{C}_{1}$, then $\mathscr{C}_{2}$ is said to be coarser than
$\mathscr{C}_{1}$, denoted as $\mathscr{C}_{1}\leq\mathscr{C}_{2}$.
\end{definition}

In other words, there exists $C^{\ast}\in \mathscr{C}_{2}$ such that
$C^{\ast}(x)\leq C(x)$ for each $C\in  \mathscr{C}_{1}$ and $x\in U$
if $\mathscr{C}_{1}\leq \mathscr{C}_{2}$.

\begin{example}
\begin{upshape}
Let $U=\{x_{1}, x_{2}, x_{3}, x_{4}\}$, $\mathscr{C}_{1}=\{C_{21},
C_{22},$ $ C_{23}\}$, and $\mathscr{C}_{2}=\{C_{40},C_{41}\}$, where
$C_{40}=\frac{0.2}{x_{1}}+\frac{0.4}{x_{2}}
+\frac{0.5}{x_{3}}+\frac{0}{x_{4}}$ and
$C_{2}=\frac{0.3}{x_{1}}+\frac{0}{x_{2}}
+\frac{0.6}{x_{3}}+\frac{0.5}{x_{4}}$. It is obvious that
$\mathscr{C}_{2}$ is coarser than $\mathscr{C}_{1}$.
\end{upshape}
\end{example}

\begin{proposition}
Let $U$ be a non-empty universe of discourse, and $\mathscr{C}_{1},
\mathscr{C}_{2}, \mathscr{C}_{3}\in C(U)$. Then

$(1)$ $\mathscr{C}_{1}\cup\mathscr{C}_{1}= \mathscr{C}_{1};$

$(2)$ $\mathscr{C}_{1}\cap\mathscr{C}_{1}\leq\mathscr{C}_{1};$

$(3)$ $\mathscr{C}_{1}\cup\mathscr{C}_{2}=
\mathscr{C}_{2}\cup\mathscr{C}_{1};$

$(4)$ $\mathscr{C}_{1}\cap\mathscr{C}_{2}=
\mathscr{C}_{2}\cap\mathscr{C}_{1};$

$(5)$ $(\mathscr{C}_{1}\cup\mathscr{C}_{2})\cup\mathscr{C}_{3}=
\mathscr{C}_{1}\cup(\mathscr{C}_{2}\cup\mathscr{C}_{3});$

$(6)$ $(\mathscr{C}_{1}\cap\mathscr{C}_{2})\cap\mathscr{C}_{3}=
\mathscr{C}_{1}\cap(\mathscr{C}_{2}\cap\mathscr{C}_{3});$

$(7)$ $\mathscr{C}_{1}\cup(\mathscr{C}_{1}\cap\mathscr{C}_{2})\leq
\mathscr{C}_{1};$

$(8)$ $\mathscr{C}_{1}\cap(\mathscr{C}_{1}\cup\mathscr{C}_{2})\leq
\mathscr{C}_{1}.$
\end{proposition}

\noindent\textbf{Proof.} Straightforward from Definitions 3.47, 3.50
and 3.53. $\Box$

\begin{proposition}
Let $U$ be a non-empty universe of discourse, and $C(U)$ the set of
all fuzzy coverings of $U$. Then $(C(U), \cap,\cup)$ is a lattice.
\end{proposition}

\noindent\textbf{Proof.} Given any $\mathscr{C}_{1},
\mathscr{C}_{2}\in C(U)$, it is obvious that $\mathscr{C}_{1}\cup
\mathscr{C}_{2}\in C(U)$ and $\mathscr{C}_{1}\cap \mathscr{C}_{2}\in
C(U)$. Therefore, $(C(U), \cap,\cup)$ is a lattice. $\Box$

We notice that
$\{\frac{1}{x_{1}}+\frac{1}{x_{2}}+...+\frac{1}{x_{n}}\}$ is the
greatest element of $(C(U), \cap,\cup)$, but $(C(U), \cap,\cup)$ is
not a complete lattice necessarily, which is illustrated by the
following example.

\begin{example}
\begin{upshape}
Let $U=\{x_{1}, x_{2}, x_{3}\}$, $C_{0}(U)=\{\mathscr{C}_{1},
\mathscr{C}_{2}, ...,\mathscr{C}_{n},...\}\subseteq C(U)$, and
$\mathscr{C}_{n}=\{\frac{\frac{1}{n}}{x_{1}}
+\frac{\frac{1}{n}}{x_{2}}+\frac{\frac{1}{n}}{x_{3}}\}$. By
Definition 3.50, it follows that $\bigcap C_{0}(U)=\{\frac{0}{x_{1}}
+\frac{0}{x_{2}}+\frac{0}{x_{3}}\}$. It is obvious that $\bigcap
C_{0}(U)\notin C(U)$. Consequently, $(C(U),\leq)$ is not an
intersection structure.
\end{upshape}
\end{example}

\begin{proposition}
Let $U$ be a non-empty universe of discourse. Then $(U, C(U)\cup
\{\emptyset\})$ is a topological space.
\end{proposition}

\noindent\textbf{Proof.} Straightforward from Definitions 3.47 and
3.50. $\Box$

At the end of this subsection, we provide two roughness measures of
fuzzy sets as follows.

\begin{definition}
Let $(U, \mathscr{C})$ be a fuzzy covering approximation space, and
$X\subseteq U$. Then the roughness measure $\mu_{\mathscr{C}}(X)$
regarding $\mathscr{C}$ is defined as
\begin{eqnarray*}
\mu_{\mathscr{C}}(X)=1-\frac{|\underline{X}_{\mathscr{C}}|}
{|\overline{X}_{\mathscr{C}}|},
\end{eqnarray*}
where $|\underline{X}_{\mathscr{C}}|=\sum_{x\in U}
\underline{X}_{\mathscr{C}}(x)$ and
$|\overline{X}_{\mathscr{C}}|=\sum_{x\in U}
\overline{X}_{\mathscr{C}}(x)$.
\end{definition}

\begin{definition}
Let $(U, \mathscr{C})$ be a fuzzy covering approximation space, and
$X\subseteq U$. Then the $\alpha\beta-$roughness measure
$\mu^{\alpha\beta}_{\mathscr{C}}(X)$ with respect to $\mathscr{C}$
is defined as
\begin{eqnarray*}
\mu^{\alpha\beta}_{\mathscr{C}}(X)=1-\frac{|\underline{X}^{\alpha}_{\mathscr{C}}|}
{|\overline{X}^{\beta}_{\mathscr{C}}|},
\end{eqnarray*}
where
$\underline{X}^{\alpha}_{\mathscr{C}}=\{x|\underline{X}_{\mathscr{C}}(x)>\alpha,
x\in U\}$,
$\overline{X}^{\beta}_{\mathscr{C}}=\{x|\overline{X}_{\mathscr{C}}(x)>\beta,
x\in U\}$  and $|\cdot|$ means the cardinality of the set.
\end{definition}

An example is employed to illustrate Definitions 3.59 and 3.60 as
follows.

\begin{example}
\begin{upshape}
Let $U=\{x_{1}, x_{2}, x_{3}, x_{4}\}$, $\mathscr{C}=\{C_{21},
C_{22}, C_{23}\}$, and $X=\frac{0.2}{x_{1}}
+\frac{0.5}{x_{2}}+\frac{0.6}{x_{3}}+\frac{0.1}{x_{4}}$. According
to Definition 3.8, we have that
$\underline{X}_{\mathscr{C}}=\frac{0.2}{x_{1}}
+\frac{0.4}{x_{2}}+\frac{0.5}{x_{3}}+\frac{0}{x_{4}}$ and
$\overline{X}_{\mathscr{C}}=\frac{0.2}{x_{1}}
+\frac{0.4}{x_{2}}+\frac{0.5}{x_{3}}+\frac{0.5}{x_{4}}$. It follows
that
$\mu_{\mathscr{C}}(X)=1-\frac{0.2+0.4+0.5}{0.2+0.4+0.5+0.5}=0.3125$.
Furthermore, it is obvious that
$\underline{X}^{\alpha}_{\mathscr{C}}=\{x_{3}\}$ and
$\overline{X}^{\beta}_{\mathscr{C}}=\{x_{2}, x_{3}, x_{4}\}$ by
taking $\alpha=0.4$ and $\beta=0.2$. Subsequently, it follows that
$\mu^{\alpha\beta}_{\mathscr{C}}(X)=1-\frac{|\{x_{3}\}|}{|\{x_{2},
x_{3}, x_{4}\}|}=\frac{2}{3}$.
\end{upshape}
\end{example}

\section{Consistent functions for fuzzy covering information systems}

In \cite{Wang2}, Wang et al. proposed the concept of consistent
functions for attribute reductions of covering information systems.
But so far we have not seen the similar work on fuzzy covering
information systems. In this section, we introduce the concepts of
consistent functions, the fuzzy covering mappings and inverse fuzzy
covering mappings based on fuzzy coverings and examine their basic
properties. Additionally, several examples are employed to
illustrate our proposed notions.

As a generalization of the concept of consistent functions given in
Definition 2.6, we introduce the notion of consistent functions for
constructing attribute reducts of fuzzy covering information
systems.

\begin{definition}
Let $f$ be a mapping from $U_{1}$ to $U_{2}$,
$\mathscr{C}$=$\{C_{1}, C_{2},...,C_{N}\}$ a fuzzy covering of
$U_{1}$, and $[x]$ the block of $U_{1}/IND(f)$ which contains $x$,
where $U_{1}/IND(f)$ stands for the blocks of partition of $U_{1}$
by an equivalence relation $IND(f)$ based on $f$. If
$C_{i}(y)=C_{i}(z)$ $(1\leq i\leq N)$ for any $y, z\in [x]$, then
$f$ is called a consistent function with respect to $\mathscr{C}$.
\end{definition}

Unless stated otherwise, we take the equivalence relation
$IND(f)=\{(x, y)|f(x)=f(y), x, y\in U_{1}\}$ and $[x]=\{y\in
U_{1}|f(x)=f(y), x,y\in U_{1}\}$ when applying Definition 4.1 in
this work.  Particularly, it is clear that our proposed function is
the same as the consistent function in \cite{Wang2} when the
membership degree for any $ x\in U_{1}$ has its value only from the
set $\{0,1\}$. Thereby, the proposed model can be viewed as an
extension of that given in \cite{Wang2}.

An example is employed to illustrate the concept of consistent
functions in the following.

\begin{example}
\begin{upshape}
Consider the fuzzy covering approximation space
$(U_{1},\mathscr{C}_{1})$ in Example 3.5. Then, we take
$U_{2}=\{y_{1}, y_{2}\}$ and define a mapping $f:
U_{1}\longrightarrow U_{2}$ as
$$f(x_{1})=f(x_{3}) = y_{1}; f(x_{2}) = f(x_{4}) = y_{2}.$$
Obviously, $f$ is a consistent function with respect to
$\mathscr{C}_{1}$.
\end{upshape}
\end{example}

Now we investigate the relationship between Definitions 2.7 and 4.1.
If $\{R_{x}|x\in U_{1}\}$ is a fuzzy covering of $U_{1}$,  where
$R_{x}(y)=R(x,y)$ for any $x, y\in U_{1}$, then we can express
Definition 2.7 as follows: let $U_{1}$ and $U_{2}$ be two universes,
$f$ a mapping from $U_{1}$ to $U_{2}$, $R\in \mathscr{F}(U_{1}\times
U_{1})$, and $[x]_{f}=\{y\in U_{1}|f(x)=f(y)\}$, $\{[x]_{f}|x\in
U_{1}\}$ a partition on $U_{1}$. For any $x,y\in U_{1}$, if
$R(x,v)=R(x,t)$ for any two pairs $(x,v),(x,t)\in
[x]_{f}\times[y]_{f}$, then $f$ is said to be consistent with
respect to $R$. Consequently, the consistent function given in
Definition 2.7 is the same as our proposed model if $\{R_{x}|x\in
U_{1}\}$ is a fuzzy covering of $U_{1}$ and $IND(f)=\{(x,
y)|f(x)=f(y), x, y\in U_{1}\}$.

In the following, we investigate some conditions under which
$\{R_{x}|x\in U_{1}\}$ is a fuzzy covering of $U_{1}$.

\begin{corollary}
Let $R$ be a fuzzy relation on $U_{1}$. Then

$(1)$ if $R$ is $\alpha$-reflexive, then $\{R_{x}|x\in U_{1}\}$ is a
fuzzy covering of $U_{1}$, where $1>\alpha>0$;

$(2)$ if $R$ is reflexive, then $\{R_{x}|x\in U_{1}\}$ is a fuzzy
covering of $U_{1}$;

$(3)$ if $R$ is a fuzzy similarity, then $\{R_{x}|x\in U_{1}\}$ is a
fuzzy covering of $U_{1}$;

$(4)$ if $R$ is a fuzzy equivalence, then $\{R_{x}|x\in U_{1}\}$ is
a fuzzy covering of $U_{1}$.
\end{corollary}

\noindent\textbf{Proof.} $(1)$ If $R$ is $\alpha$-reflexive, then
$R_{x}(x)\geq \alpha$ for any $x\in U_{1}$. It follows that
$(\bigcup\{R_{x}|x\in U_{1}\})(y)\geq\alpha$ for any $y\in U_{1}$.
Therefore, $\{R_{x}|x\in U_{1}\}$ is a fuzzy covering of $U_{1}$.

$(2)$ If $R$ is reflexive, then $R_{x}(x)=1$ for any $x\in U_{1}$.
It implies that $\bigcup\{R_{x}|x\in U_{1}\}=U_{1}$. Therefore,
$\{R_{x}|x\in U_{1}\}$ is a fuzzy covering of $U_{1}$.

$(3),(4)$ The proof is similar to that in Corollary 4.3(2). $\Box$

Additionally, we can construct a fuzzy relation by a fuzzy covering.

\begin{corollary}
Let $f$ be a mapping from $U_{1}$ to $U_{2}$,
$\mathscr{C}$=$\{C_{1}, C_{2},...,C_{N}\}$ a fuzzy covering of
$U_{1}$, $R_{x}=C_{x}$ for any $x\in U_{1}$, and
$\alpha=min\{C_{x}(x)|x\in U_{1}\}$. Then

$(1)$ $R$  is a $\alpha$-reflexive relation;

$(2)$ $R$ is symmetric if $C_{x}(x)=C_{y}(y)$ for any $x, y\in
U_{1}$;

$(3)$ $R$ is transitive;

$(4)$ $R$ is a fuzzy equivalence relation if $\alpha=1$.
\end{corollary}

\noindent\textbf{Proof.} Straightforward from Definition 3.1. $\Box$

By Corollaries 4.3 and 4.4, it is clear that there exists a
relationship between a fuzzy relation and a fuzzy covering. Since
both Wang's model\cite{Wang2} and our proposed function are based on
a fuzzy relation and a fuzzy covering, respectively, by Corollaries
4.3 and 4.4, we can establish the relationship between Definitions
2.7 and 4.1.

By means of Zadeh's extension principle, we propose the concepts of
the fuzzy covering mapping and inverse fuzzy covering mapping.
\begin{definition}Let $S_{1}=(U_{1}, \mathscr{C}_{1})$ and
$S_{2}=(U_{2},\mathscr{C}_{2})$ be fuzzy covering approximation
spaces, and $f$ a surjection from $U_{1}$ to $U_{2}$, $f$ induces a
mapping from $\mathscr{C}_{1} \text{ to } \mathscr{C}_{2}$ and a
mapping from $\mathscr{C}_{2} \text{ to }\mathscr{C}_{1}$, that is

$\hat{f}: \mathscr{C}_{1}\longrightarrow \mathscr{C}_{2},
C\mid\rightarrow \hat{f}(C)\in \mathscr{C}_{2}, \forall C\in
\mathscr{C}_{1};$

$\hat{f}(C)(y)=\left\{
\begin{array}{ccc}
\bigvee_{x\in f^{-1}(y)}C(x),&{\rm }& f^{-1}(y)\neq\emptyset;\\
0 ,&{\rm }& f^{-1}(y)=\emptyset;
\end{array}
\right.$

$\hat{f}^{-1}: \mathscr{C}_{2}\longrightarrow \mathscr{C}_{1},
T\mid\rightarrow \hat{f}^{-1}(T)\in \mathscr{C}_{1}, \forall T\in
\mathscr{C}_{2};$

$ f^{-1}(T)(x)=T(f(x)), x\in U_{1}.$\\ Then $\hat{f}$ and
$\hat{f}^{-1}$ are called the fuzzy covering mapping and the inverse
fuzzy covering mapping induced by $f$, respectively. In convenience,
we denote $\hat{f}$ and $\hat{f}^{-1}$ as $f$ and $f^{-1}$,
respectively.\end{definition}

By Definition 4.5, we observe that $\hat{f}$ and $\hat{f}^{-1}$ will
be reduced to Definition 4.1 in \cite{Wang2} if the membership
degree takes values from the set $\{0,1\}$. The following theorem
discusses the problem of fuzzy set operations under a consistent
function $f$.

\begin{theorem}
Let $f$ be a mapping from $U_{1}$ to $U_{2}$,
$\mathscr{C}$=$\{C_{1}, C_{2},...,C_{N}\}$ a fuzzy covering of
$U_{1}$, and $C_{i}, C_{j} \in \mathscr{C}$. Then

$(1)$ $f(C_{i}\cap C_{j})\subseteq f(C_{i})\cap f(C_{j})$;

$(2)$ $f(C_{i}\cup C_{j})= f(C_{i})\cup f(C_{j})$;

$(3)$  If  $f$ is a consistent function with respect to
$\mathscr{C}$, then $f(C_{i}\cap C_{j})= f(C_{i})\cap f(C_{j})$.

\end{theorem}

\noindent\textbf{Proof.} (1) By Definition 4.5, we obtain that
$f(C_{i}\cap C_{j})(f(x))=0$ when  $(C_{i}\cap C_{j})(x)=0$ for $
x\in U_{1}$. Moreover, by Definition 4.5, it follows that
$f(C_{i}\cap C_{j})(y)=\bigvee_{x'\in f^{-1}(y)}(C_{i}\cap
C_{j})(x')=\bigvee_{x'\in f^{-1}(y)}(C_{i}(x')\wedge C_{j}(x'))\leq
\bigvee_{x'\in f^{-1}(y)}C_{i}(x') \wedge \bigvee_{x'\in
f^{-1}(y)}C_{j}(x')=(f(C_{i}) \cap f(C_{j}))(y)$. Consequently,
$f(C_{i}\cap C_{j})\subseteq f(C_{i})\cap f(C_{j})$.

(2) According to Definition 4.5, we have that  $f(C_{i}\cup
C_{j})(f(x))=0$ when  $(C_{i}\cup C_{j})(x)=0$ for $ x\in U_{1}$.
Furthermore, by Definition 4.5, it follows  that $f(C_{i}\cup
C_{j})(y)=\bigvee_{x'\in f^{-1}(y)}(C_{i}\cup
C_{j})(x')=\bigvee_{x'\in f^{-1}(y)}(C_{i}(x')\vee C_{j}(x'))=
\bigvee_{x'\in f^{-1}(y)}C_{i}(x') \vee \bigvee_{x'\in
f^{-1}(y)}C_{j}(x')=(f(C_{i}) \cup f(C_{j}))(y)$. Therefore,
 $f(C_{i}\cup C_{j})= f(C_{i})\cup f(C_{j})$.

(3) By Theorem 4.6(1), it is obvious that $f(C_{i}\cap
C_{j})\subseteq f(C_{i})\cap f(C_{j})$. So we only need to prove
that $f(C_{i})\cap f(C_{j})\subseteq f(C_{i}\cap C_{j}). $ Suppose
that $y\in U_{2}$, there exists $ x\in U_{1}$  such that $f(x)=y$.
Based on Definitions 4.1 and 4.5, we have that $(f(C_{i})\cap
f(C_{j}))(y)=\bigvee_{x'\in f^{-1}(y)}C_{i}(x') \wedge
\bigvee_{x'\in f^{-1}(y)}C_{j}(x')=C_{i}(x')\wedge
C_{j}(x')\subseteq\bigvee_{x'\in f^{-1}(y)}(C_{i}(x')\wedge
C_{j}(x'))=f(C_{i}\cap C_{j})(y)$. Thereby, $f(C_{i}\cap C_{j})=
f(C_{i})\cap f(C_{j})$.        $\Box$

Theorem 4.6 shows that the mapping $f$ preserves some fuzzy set
operations, especially it preserves the intersection operation of
fuzzy sets if $f$ is consistent.

To illustrate Theorem 4.6, we give an example below.

\begin{example}\begin{upshape}
Consider $S=(U_{1}, \mathscr{C}_{1})$ in Example 3.5 and the
consistent function $f$ in Example 4.2. Then we observe that
$f(C_{1}\cap C_{2})= f(C_{1})\cap f(C_{2})$, $f(C_{1}\cap C_{3})=
f(C_{1})\cap f(C_{3})$ and $f(C_{2}\cap C_{3})= f(C_{2})\cap
f(C_{3})$.
\end{upshape}\end{example}

By Theorem 4.6, we obtain the following corollary.

\begin{corollary}
Let $f$ be a mapping from $U_{1}$ to $U_{2}$, and
$\mathscr{C}$=$\{C_{1}, C_{2},...,C_{N}\}$ a fuzzy covering of
$U_{1}$. If $f$ is a consistent function with respect to
$\mathscr{C}$, then $f(\bigcap _{i=1}^{N}C_{i})= \bigcap
_{i=1}^{N}f(C_{i})$.
\end{corollary}

Subsequently, we investigate the properties of the inverse mapping
of a consistent function.

\begin{theorem}
Let $f$ be a mapping from $U_{1}$ to $U_{2}$,
$\mathscr{C}$=$\{C_{1}, C_{2},...,C_{N}\}$ a fuzzy covering of
$U_{1}$, and $C_{i}\in \mathscr{C}$. Then

$(1)$ $ C_{i}\subseteq f^{-1}(f(C_{i}))$;

$(2)$  If $f$ is a consistent function with respect to
$\mathscr{C}$, then $f^{-1}(f(C_{i}))= C_{i}$.
\end{theorem}

\noindent\textbf{Proof.} (1) According to Definition 4.5,  we have
that $f^{-1}(f(C_{i}))(x)=f(C_{i})(f(x))$. Taking $y=f(x)$, it
follows that $f(C_{i})(f(x))=f(C_{i})(y)=\bigvee_{x'\in
f^{-1}(y)}C_{i}(x')\geq C_{i}(x).$ Therefore, $ C_{i} \subseteq
f^{-1}(f(C_{i}))$.

(2) By Definition 4.5,  we see that
$f^{-1}(f(C_{i}(x)))=f(C_{i})(f(x))$. Assume that  $y=f(x)$, it
follows that $f(C_{i})(f(x))=f(C_{i})(y)=\bigvee_{x'\in
f^{-1}(y)}C_{i}(x').$ According to Definitions 4.1 and 4.5,  it
implies that $C_{i}(x')=C_{i}(x)$ for any $ x'\in f^{-1}(y)$.
Consequently, $\bigvee_{x'\in f^{-1}(y)}C_{i}(x')=C_{i}(x)$. Hence,
$f^{-1}(f(C_{i}))(x)=C_{i}(x).$ Thereby, $ C_{i} =
f^{-1}(f(C_{i}))$.        $\Box$

We give an example to illustrate Theorem 4.9 in the following.

\begin{example}
\begin{upshape}
Consider $S=(U_{1}, \mathscr{C}_{1})$ in Example 3.5 and the
consistent function $f$ in Example 4.2. Then we see that
$f^{-1}(f(C_{1}))(x_{i})=C_{1}(x_{i}),$
$f^{-1}(f(C_{2}))(x_{i})=C_{2}(x_{i}),$ and
$f^{-1}(f(C_{3}))(x_{i})=C_{3}(x_{i}), i=1,2,3,4.$ Therefore,
$f^{-1}(f(C_{1}))=C_{1},$ $f^{-1}(f(C_{2}))=C_{2},$ and
$f^{-1}(f(C_{3}))=C_{3}.$
\end{upshape}
\end{example}

By Theorem 4.9, we have the following corollary.

\begin{corollary}
Let $f$ be a mapping from $U_{1}$ to $U_{2}$, and
$\mathscr{C}$=$\{C_{1}, C_{2},...,C_{N}\}$ a fuzzy covering of
$U_{1}$. If  $f$ is a consistent function with respect to
$\mathscr{C}$, then $f^{-1}(f(\bigcap _{i=1}^{N}C_{i}))= \bigcap
_{i=1}^{N}C_{i}$.
\end{corollary}

We also explore the properties of a consistent function on a family
of fuzzy coverings.

\begin{theorem}
Let $f$ be a mapping from $U_{1}$ to $U_{2}$, and ${\mathscr{C}_{1},
\mathscr{C}_{2}} \in C(U_{1})$. If $f$ is a consistent function with
respect to $\mathscr{C}_{1}$ and $\mathscr{C}_{2}$,  respectively,
then $f$ is consistent with respect to $\mathscr{C}_{1}\bigcap
\mathscr{C}_{2}$.
\end{theorem}

\noindent\textbf{Proof.} Based on  Definition 4.1, we have that
$C_{ix}(y)=C_{ix}(z)$ for any $y, z \in [x]$, where $i=1,2$. It
follows that $C_{1x}(y)\wedge C_{2x}(y)=C_{1x}(z)\wedge C_{2x}(z)$
for any $y, z \in [x]$. Hence, $(C_{1x}\cap C_{2x})(y)=(C_{1x}\cap
C_{2x})(z)$ for any  $y, z \in [x]$. Therefore, $f$ is consistent
with respect to $\mathscr{C}_{1}\bigcap \mathscr{C}_{2}$.
$\Box$

The following example is employed to illustrate Theorem 4.12.

\begin{example}
\begin{upshape}
Consider $S=(U_{1},\Delta)$ in Example 3.7. We take
$U_{2}=\{y_{1},y_{2}\}$ and define a mapping $f: U_{1}\rightarrow
U_{2}$ as follows:
$$f(x_{1})=f(x_{2})=y_{1},f(x_{3})=f(x_{4})=y_{2}.$$ It is obvious
that $f$ is a consistent function with respect to $\mathscr{C}_{1}$
and $ \mathscr{C}_{2}$, respectively. By Definition 4.1, we observe
that $f$ is consistent with respect to $\mathscr{C}_{1}\cap
\mathscr{C}_{2}$.
\end{upshape}
\end{example}

Based on Theorem 4.12, we obtain the following corollary.

\begin{corollary}
Let  $\mathscr{C}_{1}, \mathscr{C}_{2},..., \mathscr{C}_{m} \in
C(U_{1})$, and $f$ a mapping from $U_{1}$ to $U_{2}$. If $f$ is a
consistent function with respect to any $\mathscr{C}_{i}$ $(1\leq
i\leq m)$, then $f$ is consistent with respect to $\bigcap
_{i=1}^{m}\mathscr{C}_{i}$.
\end{corollary}

Now, we introduce two concepts for fuzzy covering approximation
spaces.

\begin{definition}
Let $f$ be a mapping from $U_{1}$ to $U_{2}$,
$\mathscr{C}_{1}=\{C_{11},C_{12},...,C_{1N}\}\in C(U_{1})$, and
$\mathscr{C}_{2}=\{T_{21},T_{22},...,T_{2M}\}\in C(U_{2})$. Then
$f(\mathscr{C}_{1})$ and $f(\mathscr{C}_{2})$ are defined by
\begin{eqnarray*}
f(\mathscr{C}_{1})&=&\{f(C_{1i}), C_{1i}\in \mathscr{C}_{1}, 1\leq i \leq  N \};\\
f^{-1}(\mathscr{C}_{2})&=&\{f^{-1}(T_{2j}), T_{2j}\in
\mathscr{C}_{2}, 1\leq j \leq  M \}.
\end{eqnarray*}
\end{definition}

\begin{theorem}
Let $U$ be a non-empty universe of discourse, and $\mathscr{C}\in
C(U)$. If $f$ is a consistent function with respect to
$\mathscr{C}$, then $f^{-1}(f(\mathscr{C}))=\mathscr{C}.$
\end{theorem}

\noindent\textbf{Proof.}  By  Theorem 4.9, it follows that
$f^{-1}(f(C_{i}))=C_{i}$ for any $C_{i}\in \mathscr{C}$. Therefore,
$f^{-1}(f(\mathscr{C}))=\mathscr{C}.$        $\Box$

Obviously, Examples 3.5 and 4.2 can illustrate Theorem 4.16. Then we
get the following corollary.

\begin{corollary}
Let $\mathscr{C}_{i}\in C(U)$, and $\Delta=\{\mathscr{C}_{i}|
i=1,2,...,m\}$. If $f$ is a consistent function with respect to any
$ \mathscr{C}_{i}\in \Delta$, then
$f^{-1}(f(\bigcap\Delta))=\bigcap\Delta.$
\end{corollary}

At the end of this section, we discuss the fuzzy covering operations
under a consistent function.

\begin{theorem}
Let $f$ be a mapping from $U_{1}$ to $U_{2}$, and $\mathscr{C}_{1}$,
$\mathscr{C}_{2}\in C(U_{1})$. Then we have

$(1)$ $f(\mathscr{C}_{1}\bigcap \mathscr{C}_{2})\subseteq
f(\mathscr{C}_{1})\bigcap f(\mathscr{C}_{2})$;

$(2)$ If $f$ is a consistent function with respect to
$\mathscr{C}_{1}$ and  $\mathscr{C}_{2}$, respectively, then
$f(\mathscr{C}_{1}\bigcap \mathscr{C}_{2})=
f(\mathscr{C}_{1})\bigcap f(\mathscr{C}_{2})$.
\end{theorem}

\noindent\textbf{Proof.} (1) According to Definitions  4.1 and 4.5,
it is obvious  that $f(\mathscr{C}_{1}\bigcap
\mathscr{C}_{2})\subseteq f(\mathscr{C}_{1})\bigcap
f(\mathscr{C}_{2})$.

(2) Evidently, we only need to prove that $
f(\mathscr{C}_{1})\bigcap f(\mathscr{C}_{2})\subseteq
f(\mathscr{C}_{1}\bigcap \mathscr{C}_{2})$. Assume that  $C_{x}$ is
the minimal element containing $x$ in $\mathscr{C}_{1}\bigcap
\mathscr{C}_{2}$, $C_{1x}$ is  the minimal element containing $x$ in
$Cov(\mathscr{C}_{1})$, and $C_{2x}$  is the minimal element
containing $x$ in $Cov(\mathscr{C}_{2})$  for any $  x\in U_{1}$. By
Definition 3.50, it follows that $C_{x}=C_{1x}\bigcap C_{2x}$.
According to Theorem 4.12, it implies that $f$ is a consistent
function with respect to $\mathscr{C}_{1}\bigcap \mathscr{C}_{2}$.
Consequently, we obtain that $f(C_{x})=f(C_{1x})\cap f(C_{2x})$. By
Definition 3.1, we have that $C_{x}(x)>0$ for any $ x\in U_{1}$. It
follows that $f(C_{x})(f(x))>0$. Hence, $(f(C_{1x})\cap
f(C_{2x}))(f(x))>0.$ Suppose that $f(C_{1x})\cap f(C_{2x})$ is not
the minimal subset containing $f(x)$ in $f(\mathscr{C}_{1}\bigcap
\mathscr{C}_{2})$. Then there exists $x_{0}\in U_{1}$ such that
$f(C_{ix_{0}})(f(x))>0$ and $f(C_{1x})\cap f(C_{2x})\cap
f(C_{ix_{0}}) \subset f(C_{1x})\cap f(C_{2x}),$ it means that
$(f(C_{1x})\cap f(C_{2x})\cap f(C_{ix_{0}}))(f(x))>0.$ Thereby,
there exist $u, v$ and $w$ such that $C_{1x}(u)>0, C_{2x}(v)>0,
C_{ix_{0}}(w)>0$ and $f(u)=f(v)=f(w)=f(x)$. According to Theorem
4.6, we have that $f(C_{1x})\cap f(C_{2x})=f(C_{1x}\cap
C_{2x})\subseteq f(C_{ix_{0}})$   and $f(C_{1x})\cap f(C_{2x})\cap
f(C_{ix_{0}}) = f(C_{1x})\cap f(C_{2x}),$ it implies that
$f(C_{1x})\cap f(C_{2x}) $ is the minimal subset containing $f(x)$
in $f(\mathscr{C}_{1}\bigcap \mathscr{C}_{2})$. Based on the above
statement, it follows that $ f(\mathscr{C}_{1})\bigcap
f(\mathscr{C}_{2})\subseteq f(\mathscr{C}_{1}\bigcap
\mathscr{C}_{2})$. Therefore, $ f(\mathscr{C}_{1}\bigcap
\mathscr{C}_{2})= f(\mathscr{C}_{1})\bigcap f(\mathscr{C}_{2})$.
$\Box$

Based on Theorem 4.18, we have the following corollary.

\begin{corollary}
Let  $\mathscr{C}_{1}, \mathscr{C}_{2},...,\mathscr{C}_{m}$ be fuzzy
coverings of $U_{1}$, and $f$ a mapping from $U_{1}$ to $U_{2}$. If
$f$ is a consistent function with respect to $\mathscr{C}_{1},
\mathscr{C}_{2},..., \mathscr{C}_{m}$, respectively,  then
$f(\bigcap _{i=1}^{m}\mathscr{C}_{i})= \bigcap_{i=1}^{m}
f(\mathscr{C}_{i})$.
\end{corollary}

\section{Data compressions of fuzzy covering information systems
and dynamic fuzzy covering information systems}

In this section, we further investigate data compressions of fuzzy
covering information systems and dynamic fuzzy covering information
systems.

\subsection{Data compression of fuzzy covering information systems}
In this subsection, the concepts of an induced fuzzy covering
information system and homomorphisms between fuzzy covering
information systems are introduced for data compression of the fuzzy
covering information system. Then the algorithm of constructing
attribute reducts of fuzzy covering information systems is provided.
An example is finally employed to illustrate the proposed concepts
and algorithm.

\begin{definition}
Let $f$ be a surjection from $U_{1}$ to $U_{2}$,
$\Delta_{1}$=$\{\mathscr{C}_{1},
\mathscr{C}_{2},...,\mathscr{C}_{m}\}$ a family of fuzzy coverings
of $U_{1}$, and $f(\Delta_{1})$=$\{f(\mathscr{C}_{1}),
f(\mathscr{C}_{2}),...,f(\mathscr{C}_{m})\}$. Then $(U_{1},
\Delta_{1})$ is referred to as a fuzzy covering information system
and  $(U_{2}, f(\Delta_{1}))$ is called the $f$-induced fuzzy
covering information system of $(U_{1}, \Delta_{1})$.
\end{definition}

Definition 5.1 shows that we can induce a new fuzzy covering
information system under a surjection.

Based on Definitions 4.1 and 5.1, we propose the notion of a
homomorphism between two fuzzy covering information systems.

\begin{definition}
Let $f$ be a surjection from $U_{1}$ to $U_{2}$,
$\Delta_{1}$=$\{\mathscr{C}_{1},
\mathscr{C}_{2},...,\mathscr{C}_{m}\}$ a family of fuzzy coverings
of $U_{1}$, and $f(\Delta_{1})$=$\{f(\mathscr{C}_{1}),
f(\mathscr{C}_{2}),...,f(\mathscr{C}_{m})\}$. If $f$ is consistent
with respect to any $\mathscr{C}_{i}\in \Delta_{1}$ $(1\leq i\leq
m)$ on $U_{1}$, then $f$ is called a homomorphism from $(U_{1},
\Delta_{1})$ to $(U_{2}, f(\Delta_{1}))$.
\end{definition}

We provide the concept of reducts of fuzzy covering information
systems in the following.

\begin{definition}
Let $(U_{1},\Delta_{1}) $ be a fuzzy covering information system,
and $\mathscr{C}_{i}\in \Delta_{1}$ $(1\leq i\leq m)$. If
$\bigcap\{\Delta_{1}-\mathscr{C}_{i}\}=\bigcap\Delta_{1}\ $, then
$\mathscr{C}_{i}$ is called superfluous. Otherwise,
$\mathscr{C}_{i}$ is called indispensable. The collection of all
indispensable elements in $\Delta_{1}$, denoted as
Core($\Delta_{1}$), is called the core of $\Delta_{1}$. $P\subseteq
\Delta_{1}$ is called a reduct of $\Delta_{1}$ if $P$ satisfies:
$\bigcap P=\bigcap \Delta_{1}$ and $\bigcap \{P-\mathscr{C}\}\neq
\bigcap \Delta_{1}$ for any $ \mathscr{C}\in P.$
\end{definition}

Now we present the following theorem which shows that the reducts of
fuzzy covering information systems can be preserved under a
homomorphism.

\begin{theorem}
Let $f$ be a surjection from $U_{1}$ to $U_{2}$,
$\Delta_{1}$=$\{\mathscr{C}_{1},
\mathscr{C}_{2},...,\mathscr{C}_{m}\}$ a family of fuzzy coverings
of $U_{1}$, and $f(\Delta_{1})$=$\{f(\mathscr{C}_{1}),
f(\mathscr{C}_{2}),...,f(\mathscr{C}_{m})\}$. If $f$ is a
homomorphism from $(U_{1}, \Delta_{1})$ to $(U_{2}, f(\Delta_{1}))$,
then $P\subseteq \Delta_{1}$ is a reduct of $\Delta_{1}$ if and only
if $f(P)$ is a reduct of $f(\Delta_{1})$.
\end{theorem}

\noindent\textbf{Proof.} Suppose $P$ is a reduct of $\Delta_{1}$. It
follows that $\bigcap P=\bigcap \Delta_{1}$. Hence, $f(\bigcap
P)=f(\bigcap \Delta_{1})$. Then we obtain that $\bigcap f(P)=\bigcap
f(\Delta_{1})$ since $f$ is a homomorphism from $(U_{1},
\Delta_{1})$ to $(U_{2}, f(\Delta_{1}))$. Assume that there exists
$\mathscr{C}\in P$ such that $\bigcap (f(P)-f(\mathscr{C}))=\bigcap
f(P)$. It implies that $\bigcap (f(P)-f(\mathscr{C}))=\bigcap
f(P-\mathscr{C})$. Hence, we see that $\bigcap f(\Delta_{1})=\bigcap
f(P-\mathscr{C})$. It follows that $f^{-1}(\bigcap
f(\Delta_{1}))=f^{-1}(\bigcap f(P-\mathscr{C}))$. We obtain that
$\bigcap \Delta_{1}=\bigcap (P-\mathscr{C})$, which contradicts that
$P$ is a reduct of $\Delta_{1}$. So $f(P)$ is a reduct of
$f(\Delta_{1})$.

On the other hand, we assume that $f(P)$ is a reduct of
$f(\Delta_{1})$. It follows that $\bigcap f(\Delta_{1})=\bigcap
f(P)$. Since $f$ is a homomorphism from $(U_{1}, \Delta_{1})$ to
$(U_{2}, f(\Delta_{1}))$, we obtain that $
f(\bigcap\Delta_{1})=f(\bigcap P)$. It implies that
$\bigcap\Delta_{1}=\bigcap P$. Assume that there exists
$\mathscr{C}\in P$ satisfying $\bigcap\Delta_{1}=\bigcap
(P-\mathscr{C})$, it follows that $f(\bigcap\Delta_{1})=f(\bigcap
(P-\mathscr{C}))$. Obviously, $ \bigcap f(\Delta_{1})=\bigcap f
(P-\mathscr{C})=\bigcap (f(P)-f(\mathscr{C}))$, which is a
contradiction. Therefore, $P\subseteq \Delta_{1}$ is a reduct of
$\Delta_{1}$.       $\Box$

By Theorem 5.4, we obtain the following corollary.

\begin{corollary}
Let $f$ be a surjection from $U_{1}$ to $U_{2}$,
$\Delta_{1}$=$\{\mathscr{C}_{1},
\mathscr{C}_{2},...,\mathscr{C}_{m}\}$ a family of fuzzy coverings
of $U_{1}$, and $f(\Delta_{1})$=$\{f(\mathscr{C}_{1}),
f(\mathscr{C}_{2}),...,f(\mathscr{C}_{m})\}$. If $f$ is a
homomorphism from $(U_{1}, \Delta_{1})$ to $(U_{2}, f(\Delta_{1}))$,
then

$(1)$ $\mathscr{C}$ is indispensable in $\Delta_{1}$ if and only if
$f(\mathscr{C})$ is indispensable in $f(\Delta_{1})$;

$(2)$ $\mathscr{C}$ is superfluous in $\Delta_{1}$ if and only if
$f(\mathscr{C})$ is superfluous in $f(\Delta_{1})$;

$(3)$ The image of the core of $\Delta_{1}$ is the core of
$f(\Delta_{1})$, and the inverse image of the core of
$f(\Delta_{1})$ is the core of the original image.
\end{corollary}

\noindent\textbf{Proof.} Straightforward from Definition 5.3 and
Theorem 5.4.       $\Box$

From Corollary 5.5, we see that the attribute reductions of the
original fuzzy covering information system and image system are
equivalent to each other under the condition of a homomorphism.

\begin{definition}
Let $(U_{1}, \mathscr{C}_{1})$ be a fuzzy covering approximation
space, the equivalence relation
$R_{\mathscr{C}_{1}}=\{(x,y)|C_{x}=C_{y}, x,y\in U_{1}\}$, and
$U_{1}/R_{\mathscr{C}_{1}}=\{R_{\mathscr{C}_{1}}(x)|x\in U_{1}\}$.
Then $U_{1}/R_{\mathscr{C}_{1}}$ is called the partition based on
$\mathscr{C}_{1}$.
\end{definition}

For the sake of convenience, we denote $U_{1}/R_{\mathscr{C}_{1}}$
as $U_{1}/\mathscr{C}_{1}$ simply.

Following, we employ Table 2 to show the partition based on each
fuzzy covering for the fuzzy covering information system $(U_{1},
\Delta_{1})$, where $P_{ix_{j}}$ stands for the block containing
$x_{j}$ in the partition $U_{1}/R_{\mathscr{C}_{i}}$. It is easy to
see that $P_{\Delta_{1} x_{j}}=\bigcap_{1\leq i\leq m}P_{ix_{j}}$,
where $P_{\Delta_{1} x_{j}}$ denotes the block containing $x_{j}$ in
the partition based on $\Delta_{1}$.

Subsequently, we propose the main feature of the algorithm to
construct attribute reducts of fuzzy covering information systems.
It shows how to construct a homomorphism and compress a large-scale
information system into a small one under the condition of the
homomorphism.

\begin{algorithm}
Let $U_{1}=\{x_{1},...,x_{n}\}$, and
$\Delta_{1}$=$\{\mathscr{C}_{1}, \mathscr{C}_{2},$
$...,\mathscr{C}_{m}\}$ a family of fuzzy coverings of  $U_{1}$.
Then

Step 1. Input the fuzzy covering information system $(U_{1},
\Delta_{1})$;

Step 2. Computing the partition $U_{1}/\mathscr{C}_{i}$ $(1\leq
i\leq m )$ and obtain $U_{1}/\Delta_{1}=\{C_{i}|1\leq i\leq K\}$;

Step 3. Define $f: U_{1}\rightarrow U_{2} $ as follows:
$f(x)=y_{l}$, $x\in C_{l}$, where $1\leq l\leq K$ and
$U_{2}=\{y_{1},y_{2},...,y_{K}\}$;

Step 4. Compute $f(\Delta_{1})$=$\{f(\mathscr{C}_{1}),
f(\mathscr{C}_{2}),...,f(\mathscr{C}_{m})\}$  and obtain $(U_{2},
f(\Delta_{1}))$;

Step 5. Construct attribute reducts of $(U_{2}, f(\Delta_{1}))$ and
obtain a reduct $\{f(\mathscr{C}_{i1}),
f(\mathscr{C}_{i2}),...,f(\mathscr{C}_{ik})\};$

Step 6. Obtain a reduct $\{\mathscr{C}_{i1}, \mathscr{C}_{i2},
...,\mathscr{C}_{ik}\}$ of $(U_{1}, \Delta_{1})$ and output the
results.
\end{algorithm}

\noindent\textbf{Remark.} In Example 5.1\cite{Wang2}, Wang et al.
obtained the partition $U_{1}/\Delta_{1}$ by only computing
$\Delta_{x}$ for any $x\in U_{1}$. But we get $U_{1}/\Delta_{1}$ by
computing $U_{1}/\mathscr{C}_{i}$ for any $\mathscr{C}_{i}\in
\Delta_{1}$ in Algorithm 5.7. By using the proposed approach, we can
compress the dynamic fuzzy covering information system on the basis
of data compression of the original system with lower time
complexity, which is illustrated in Subsection 5.2.

Now, we employ a car evaluation problem to illustrate Algorithm 5.7.

\begin{example}
\begin{upshape}  Suppose that $U_{1}=\{x_{1}, x_{2},..., x_{8}\}$
is a set of eight cars, $C_{1}=\{price, structure, size,
appearance\}$ is a set of attributes. The domains of $price$,
$structure$, $size$ and $appearance$ are $\{high, middle,$ $low\}$,
$\{excellent, ordinary, poor\}$, $\{big, middle, small\}$ and
$\{beautiful, fair, ugly\}$, respectively. In this example, we do
not list their evaluation reports for simplicity. According to the
four specialists' evaluation reports, we obtain the following fuzzy
coverings of $U_{1}$ as $\Delta_{1}=\{\mathscr{C}_{price},
\mathscr{C}_{structure}, \mathscr{C}_{size},
\mathscr{C}_{appearance}\}$, $\mathscr{C}_{price},
\mathscr{C}_{structure}, \mathscr{C}_{size} $ and $
\mathscr{C}_{appearance}$ are based on $price$, $structure$, $size$
and $appearance$, respectively, where
\begin{eqnarray*}
\mathscr{C}_{price}&=&\{
\frac{1}{x_{1}}+\frac{1}{x_{2}}+\frac{0.5}{x_{3}}+\frac{1}{x_{4}}+\frac{0.5}{x_{5}}
+\frac{1}{x_{6}}+\frac{1}{x_{7}}+\frac{1}{x_{8}},
\frac{0.5}{x_{1}}+\frac{0.5}{x_{2}}+\frac{0.5}{x_{3}}+\frac{1}{x_{4}}+\frac{0.5}{x_{5}}
+\frac{0.5}{x_{6}}+\frac{1}{x_{7}}+\frac{1}{x_{8}},\\
&&
\frac{0}{x_{1}}+\frac{0}{x_{2}}+\frac{1}{x_{3}}+\frac{0.5}{x_{4}}+\frac{1}{x_{5}}
+\frac{1}{x_{6}}+\frac{0.5}{x_{7}}+\frac{0.5}{x_{8}}\};\\
\mathscr{C}_{structure}&=&\{
\frac{0}{x_{1}}+\frac{0}{x_{2}}+\frac{1}{x_{3}}+\frac{0}{x_{4}}+\frac{1}{x_{5}}
+\frac{0}{x_{6}}+\frac{0}{x_{7}}+\frac{0}{x_{8}},
\frac{1}{x_{1}}+\frac{1}{x_{2}}+\frac{0.5}{x_{3}}+\frac{1}{x_{4}}+\frac{0.5}{x_{5}}
+\frac{1}{x_{6}}+\frac{1}{x_{7}}+\frac{1}{x_{8}},\\ &&
\frac{1}{x_{1}}+\frac{1}{x_{2}}+\frac{0.5}{x_{3}}+\frac{0.5}{x_{4}}+\frac{0.5}{x_{5}}
+\frac{0}{x_{6}}+\frac{0.5}{x_{7}}+\frac{0.5}{x_{8}}\};\\
\mathscr{C}_{size}&=&\{
\frac{1}{x_{1}}+\frac{1}{x_{2}}+\frac{1}{x_{3}}+\frac{0}{x_{4}}+\frac{1}{x_{5}}
+\frac{1}{x_{6}}+\frac{0}{x_{7}}+\frac{0}{x_{8}},
\frac{0.5}{x_{1}}+\frac{0.5}{x_{2}}+\frac{1}{x_{3}}+\frac{0.5}{x_{4}}+\frac{1}{x_{5}}
+\frac{0.5}{x_{6}}+\frac{0.5}{x_{7}}+\frac{0.5}{x_{8}},\\
&&
\frac{1}{x_{1}}+\frac{1}{x_{2}}+\frac{1}{x_{3}}+\frac{1}{x_{4}}+\frac{1}{x_{5}}
+\frac{0.5}{x_{6}}+\frac{1}{x_{7}}+\frac{1}{x_{8}}\};\\
\mathscr{C}_{appearance}&=&\{
\frac{1}{x_{1}}+\frac{1}{x_{2}}+\frac{0.5}{x_{3}}+\frac{1}{x_{4}}+\frac{0.5}{x_{5}}
+\frac{1}{x_{6}}+\frac{1}{x_{7}}+\frac{1}{x_{8}},
\frac{1}{x_{1}}+\frac{1}{x_{2}}+\frac{0.5}{x_{3}}+\frac{1}{x_{4}}+\frac{0.5}{x_{5}}
+\frac{1}{x_{6}}+\frac{1}{x_{7}}+\frac{1}{x_{8}},\\ &&
\frac{1}{x_{1}}+\frac{1}{x_{2}}+\frac{1}{x_{3}}+\frac{1}{x_{4}}
+\frac{1}{x_{5}}+\frac{0.5}{x_{6}}+\frac{1}{x_{7}}+\frac{1}{x_{8}}\}.
\end{eqnarray*}

By Definition 2.8, we see that  $(U_{1}, \Delta_{1})$ is a fuzzy
covering information system. Furthermore, according to Definitions
3.1, 3.6 and 5.6, we obtain the following results:
\begin{eqnarray*}
U_{1}/\mathscr{C}_{price}&=&\{\{x_{1},x_{2}\},\{x_{3},x_{4},x_{5},x_{6},x_{7},x_{8}\}\};\\
U_{1}/\mathscr{C}_{structure}&=&\{\{x_{1},x_{2},x_{4},x_{7},x_{8}\},\{x_{3},x_{5}\},\{x_{6}\}\};\\
U_{1}/\mathscr{C}_{size}&=&\{\{x_{1},x_{2},x_{3},x_{5},x_{6}\},\{x_{4},x_{7},x_{8}\}\};\\
U_{1}/\mathscr{C}_{appearance}&=&\{\{x_{1},x_{2},x_{3},x_{4},x_{5},x_{6},x_{7},x_{8}\}\}.
\end{eqnarray*}
The partitions
$U_{1}/\mathscr{C}_{price},U_{1}/\mathscr{C}_{structure},U_{1}/\mathscr{C}_{size}$
and $ U_{1}/\mathscr{C}_{appearance}$ are shown in Table 3. Then we
obtain that
$U_{1}/\Delta_{1}=\{\{x_{1},x_{2}\},\{x_{3},x_{5}\},\{x_{4},
x_{7},x_{8}\},\{x_{6}\}\}.$ Thus we take $U_{2}=\{y_{1}, y_{2},
y_{3}, y_{4}\}$ and define a mapping $f: U_{1}\longrightarrow U_{2}$
as follows:
$$f(x_{1})=f(x_{2}) = y_{1}; f(x_{3}) = f(x_{5}) = y_{2};
f(x_{4})=f(x_{7})= f(x_{8}) = y_{3}; f(x_{6}) = y_{4}.$$

According to the function $f$, we obtain that
$f(\Delta_{1})=\{f(\mathscr{C}_{price}), f(\mathscr{C}_{structure}),
f(\mathscr{C}_{size}), f(\mathscr{C}_{appearance})\}$, where
\begin{eqnarray*}
f(\mathscr{C}_{price})&=&\{
\frac{1}{y_{1}}+\frac{0.5}{y_{2}}+\frac{1}{y_{3}}+\frac{1}{y_{4}},
\frac{0.5}{y_{1}}+\frac{0.5}{y_{2}}+\frac{1}{y_{3}}+\frac{0.5}{y_{4}},
\frac{0}{y_{1}}+\frac{1}{y_{2}}+\frac{0.5}{y_{3}}+\frac{1}{y_{4}}\};\\
f(\mathscr{C}_{structure})&=&\{
\frac{0}{y_{1}}+\frac{1}{y_{2}}+\frac{0}{y_{3}}+\frac{0}{y_{4}},
\frac{1}{y_{1}}+\frac{0.5}{y_{2}}+\frac{1}{y_{3}}+\frac{1}{y_{4}},
\frac{1}{y_{1}}+\frac{0.5}{y_{2}}+\frac{0.5}{y_{3}}+\frac{0}{y_{4}}\};
\\
f(\mathscr{C}_{size})&=&\{
\frac{1}{y_{1}}+\frac{1}{y_{2}}+\frac{0}{y_{3}}+\frac{1}{y_{4}},
\frac{0.5}{y_{1}}+\frac{1}{y_{2}}+\frac{0.5}{y_{3}}+\frac{0.5}{y_{4}},
\frac{1}{y_{1}}+\frac{1}{y_{2}}+\frac{1}{y_{3}}+\frac{0.5}{y_{4}}\};\\
f(\mathscr{C}_{appearance})&=&\{
\frac{1}{y_{1}}+\frac{0.5}{y_{2}}+\frac{1}{y_{3}}+\frac{1}{y_{4}},
\frac{1}{y_{1}}+\frac{0.5}{y_{2}}+\frac{1}{y_{3}}+\frac{1}{y_{4}},
\frac{1}{y_{1}}+\frac{1}{y_{2}}
+\frac{1}{y_{3}}+\frac{0.5}{y_{4}}\}.
\end{eqnarray*}
\end{upshape}
\end{example}

According to Definition 5.1, we obtain the $f$-induced fuzzy
covering information system  $(U_{2}, f(\Delta_{1}))$ of $(U_{1},
\Delta_{1})$. Clearly, the size of $(U_{2}, f(\Delta_{1}))$ is
relatively smaller than that of $(U_{1}, \Delta_{1})$. Then, by
Definitions 5.1, 5.2 and 5.3, we have the following results:

(1)  $f$ is a homomorphism from $(U_{1}, \Delta_{1})$ to $(U_{2},
f(\Delta_{1}))$;

(2) $f(\mathscr{C}_{appearance})$ is superfluous in $f(\Delta_{1})$
if and only if $\mathscr{C}_{appearance}$ is superfluous in
$\Delta_{1}$;

(3) $\{f(\mathscr{C}_{price}), f(\mathscr{C}_{structure}),
f(\mathscr{C}_{size})\}$ is a reduct of $f(\Delta_{1})$ if and only
if $\{\mathscr{C}_{price}, \mathscr{C}_{structure},
\mathscr{C}_{size}\}$ is a reduct of $\Delta_{1}$.

From Example 5.8, we see that the image system $(U_{2},
f(\Delta_{1}))$ has relatively smaller size than the original system
$(U_{1}, \Delta_{1})$. But their attribute reductions are equivalent
to each other under the condition of a homomorphism.

From the practical viewpoint, it may be difficult to construct
attribute reducts of a large-scale fuzzy covering information system
directly. However, we can compress it into a relatively smaller
fuzzy covering information system under the condition of a
homomorphism and conduct the attribute reductions on the image
system. Therefore, the notion of a homomorphism may provide a more
efficient approach to dealing with large-scale fuzzy covering
information systems.

\subsection{Data compression of dynamic fuzzy covering information systems}

In Subsection 5.1, we derive a partition based on each fuzzy
covering shown in Table 2, which is useful for data compression of
dynamic fuzzy covering information systems. In the following, we
discuss how to compress two types of dynamic fuzzy covering
information systems by utilizing the compression of the original
fuzzy covering information system.

Type 1:  Adding a family of fuzzy coverings. By adding a fuzzy
covering $\mathscr{C}_{m+1}$ to the fuzzy covering information
system $(U_{1},\Delta_{1})$, we obtain the dynamic fuzzy covering
information system $(U_{1},\Delta )$, where $\Delta=\Delta_{1}\cup
\{\mathscr{C}_{m+1}\}$. There are three steps to compress the
dynamic fuzzy covering information system $(U_{1},\Delta )$.  First,
we obtain the partition $U_{1}/\mathscr{C}_{m+1}$ in the sense of
Definition 5.6 and get Table 4 by adding $U_{1}/\mathscr{C}_{m+1}$
into Table 2. Then we derive the partition $U_{1}/\Delta$ based on
$U_{1}/\mathscr{C}_{i}$ $(1\leq i\leq m+1)$. Afterwards, we define
the homomorphism $f$ based on $U_{1}/\Delta$ as Example 5.8 and
compress $(U_{1},\Delta )$ into a small-scale fuzzy covering
information system $(f(U_{1}),f(\Delta))$. Furthermore, the same
process can be applied to the dynamic fuzzy covering information
system when adding a family of fuzzy coverings.

Type 2:  Deleting a family of fuzzy coverings. We obtain the dynamic
fuzzy covering information system $(U_{1}, \Delta)$ when deleting
the fuzzy covering $\mathscr{C}_{k}\in\Delta_{1}$, where
$\Delta=\Delta_{1}-\{\mathscr{C}_{k}\}$. To compress the dynamic
fuzzy covering information system $(U_{1}, \Delta)$, we first derive
Table 5 by canceling the partition $U_{1}/\mathscr{C}_{k}$ in Table
2. Then we obtain the partition $U_{1}/\Delta$ based on
$U_{1}/\mathscr{C}_{i}$ $ (1\leq i\leq k-1,k+1\leq i\leq m)$ and
define the homomorphism $f$ as Example 5.8. Afterwards, $(U_{1},
\Delta)$ is compressed into a small-scale fuzzy covering information
system $(f(U_{1}), f(\Delta))$. Moreover, we can compress the
dynamic fuzzy covering information system when deleting a family of
fuzzy coverings using the same approach.

In practice, it may be very costly or even intractable to construct
the compression of the dynamic fuzzy covering information system as
the original fuzzy covering information system. Thus the proposed
approach based on the compression of the original fuzzy covering
information system may provide a more efficient approach to dealing
with data compression of dynamic fuzzy covering information systems.

\section{Conclusion and further research}

In this paper, we have presented some new operations on fuzzy
coverings and investigated their properties in detail. Particularly,
the lower and upper approximation operations based on fuzzy
coverings have been introduced for the fuzzy covering approximation
space. Then we have constructed a consistent function for the
communication between fuzzy covering information systems, and
pointed out that a homomorphism is a special fuzzy covering mapping
between the two fuzzy covering information systems. In addition, we
have proved that attribute reductions of the original system and
image system are equivalent to each other under the condition of a
homomorphism. We have also applied the proposed approach to
attribute reductions of fuzzy covering information systems and
dynamic fuzzy covering information systems.

In future, we will further study the fuzzy covering information
systems by extending the covering rough sets and apply the proposed
method to feature selections of fuzzy covering information systems.
Furthermore, we will discuss the data compression of dynamic
relation information systems and dynamic fuzzy relation information
systems. Especially, we will apply an incremental updating scheme to
maintain the compression dynamically and avoid unnecessary
computations by utilizing the compression of the original system.

\section*{ Acknowledgments}
We would like to thank the anonymous reviewers very much for their
helpful comments and valuable suggestions. This work is supported by
the National Natural Science Foundation of China (NO. 11071061) and
the National Basic Research Program of China (NO. 2010CB334706).

\newpage
\begin{table}[htbp]
\caption{An incomplete information system.}
 \tabcolsep0.63in
\begin{tabular}{ c c c c}
\hline $U$  & $structure$ &$color$& $price$\\ \hline
$x_{1}$ & $bad$& $good$ &$low$ \\
$x_{2}$ & $\ast$& $good$  &$high$\\
$x_{3}$ & $good$& $bad$  &$high$\\
$x_{4}$ & $bad$& $bad$  &$\ast$\\
$x_{5}$ & $good$& $\ast$ &$low$   \\
$x_{6}$ & $\ast$& $bad$ &$\ast$ \\
\hline
\end{tabular}
\end{table}

\newpage
\begin{table}[htbp]
\caption{The partitions based on each fuzzy covering
$\mathscr{C}_{i}$ $(1\leq i\leq m )$ and $\Delta_{1}$,
respectively.}
 \tabcolsep0.31in
\begin{tabular}{c c c c c c c c}
\hline $U_{1}$  &$\mathscr{C}_{1}$& $\mathscr{C}_{2}$&.&.&. &$\mathscr{C}_{m}$& $\Delta_{1}$\\
\hline
$x_{1}$ & $P_{1x_{1}}$& $P_{2x_{1}}$&.&.&. &$P_{mx_{1}}  $ & $P_{\Delta_{1} x_{1}}$\\
$x_{2}$ & $P_{1x_{2}}$& $P_{2x_{2}}$&.&.&. &$P_{mx_{2}}  $&$P_{\Delta_{1} x_{2}}$\\
$.$     & $.$        & .&.&.&. &$ . $& .\\
$.$     & $.$        & .&.&.&. &$ . $&.\\
$.$     & $.$        & .&.&.&. &$ . $ &.\\
$x_{n}$ & $P_{1x_{n}}$& $P_{2x_{2}}$&.&.&. &$P_{mx_{n}}$&$P_{\Delta_{1} x_{n}}$  \\
\hline
\end{tabular}
\end{table}

\newpage
\begin{table}[htbp]
\caption{The partitions based on $\mathscr{C}_{prize},
\mathscr{C}_{structure},
\mathscr{C}_{size},\mathscr{C}_{appearance}$ and $\Delta_{1}$,
respectively.}
 \tabcolsep0.135in
\begin{tabular}{c c c c c c}
\hline $U_{1}$ &$\mathscr{C}_{price}$& $\mathscr{C}_{structure}$ &$\mathscr{C}_{size}$& $\mathscr{C}_{appearance}$& $\Delta_{1}$\\
\hline
$x_{1}$ & $\{x_{1}, x_{2}\}$& $\{x_{1}, x_{2}, x_{4}, x_{7},x_{8}\}$& $\{x_{1}, x_{2}, x_{3}, x_{5},x_{6}\}$ & $U_{1}$&$\{x_{1}, x_{2}\}$ \\
$x_{2}$ & $\{x_{1}, x_{2}\}$& $\{x_{1}, x_{2}, x_{4}, x_{7},x_{8}\}$& $\{x_{1}, x_{2}, x_{3}, x_{5},x_{6}\}$ &$U_{1}$ &$\{x_{1}, x_{2}\}$\\
$x_{3}$ & $\{x_{3}, x_{4},x_{5},x_{6},x_{7},x_{8}\}$& $\{x_{3}, x_{5}\}$& $\{x_{1}, x_{2}, x_{3}, x_{5},x_{6}\}$ &$U_{1}$ &$\{x_{3}, x_{5}\}$ \\
$x_{4}$ & $\{x_{3}, x_{4},x_{5},x_{6},x_{7},x_{8}\}$& $\{x_{1}, x_{2}, x_{4}, x_{7},x_{8}\}$& $\{x_{4}, x_{7},x_{8}\}$  &$U_{1}$ &$\{x_{4}, x_{7},x_{8}\}$\\
$x_{5}$ & $\{x_{3}, x_{4},x_{5},x_{6},x_{7},x_{8}\}$& $\{x_{3}, x_{5}\}$& $\{x_{1}, x_{2}, x_{3}, x_{5},x_{6}\}$ & $U_{1}$&$\{x_{3}, x_{5}\}$ \\
$x_{6}$ & $\{x_{3}, x_{4},x_{5},x_{6},x_{7},x_{8}\}$& $\{x_{6}\}$& $\{x_{1}, x_{2}, x_{3}, x_{5},x_{6}\}$  &$U_{1}$  &$\{x_{6}\}$\\
$x_{7}$ & $\{x_{3}, x_{4},x_{5},x_{6},x_{7},x_{8}\}$& $\{x_{1}, x_{2}, x_{4}, x_{7},x_{8}\}$& $\{x_{4}, x_{7},x_{8}\}$ &$U_{1}$ &$\{x_{4}, x_{7},x_{8}\}$ \\
$x_{8}$ & $\{x_{3}, x_{4},x_{5},x_{6},x_{7},x_{8}\}$& $\{x_{1}, x_{2}, x_{4}, x_{7},x_{8}\}$& $\{x_{4}, x_{7},x_{8}\}$  &$U_{1}$  &$\{x_{4}, x_{7},x_{8}\}$\\
\hline
\end{tabular}
\end{table}

\newpage
\begin{table}[htbp]
\caption{The partitions based on each fuzzy covering
$\mathscr{C}_{i}$ $(1\leq i\leq m+1 )$ and $\Delta$, respectively.}
 \tabcolsep0.25in
\begin{tabular}{c c c c c c c c c}
\hline $U_{1}$  &$\mathscr{C}_{1}$& $\mathscr{C}_{2}$&.&.&.&$\mathscr{C}_{m}$& $\mathscr{C}_{m+1}$& $\Delta$\\
\hline
$x_{1}$ & $P_{1x_{1}}$& $P_{2x_{1}}$&.&.&. &$P_{mx_{1}}$&$P_{(m+1)x_{1}}$& $P_{\Delta x_{1}}$\\
$x_{2}$ & $P_{1x_{2}}$& $P_{2x_{2}}$&.&.&. &$P_{mx_{2}}$&$P_{(m+1)x_{2}}$&$P_{\Delta x_{2}}$\\
$.$     & $.$        & .&.&.&. &$ . $& .& .\\
$.$     & $.$        & .&.&.&. &$ . $&.& .\\
$.$     & $.$        & .&.&.&. &$ . $ &.& .\\
$x_{n}$ & $P_{1x_{n}}$& $P_{2x_{2}}$&.&.&. &$P_{mx_{n}}$&$P_{(m+1)x_{n}}$&$P_{\Delta x_{n}}$  \\
\hline
\end{tabular}
\end{table}

\newpage
\begin{table}[htbp]
\caption{The partitions based on each fuzzy covering
$\mathscr{C}_{i}$ $(1\leq i\leq k-1, k+1\leq i\leq m )$ and
$\Delta$, respectively.}
 \tabcolsep0.15in
\begin{tabular}{c c c c c c c c c c c c c c c}
\hline $U_{1}$  &$\mathscr{C}_{1}$&$\mathscr{C}_{2}$&.&.&.&$\mathscr{C}_{k-1}$&$\mathscr{C}_{k+1}$&.&.&.&$\mathscr{C}_{m}$& $\Delta$\\
\hline
$x_{1}$ & $P_{1x_{1}}$&$P_{2x_{1}}$&.&.&.&$P_{(k-1)x_{1}}$&$P_{(k+1)x_{1}}$&.&.&. &$P_{mx_{1}}$& $P_{\Delta x_{1}}$\\
$x_{2}$ & $P_{1x_{2}}$&$P_{2x_{2}}$&.&.&.&$P_{(k-1)x_{2}}$&$P_{(k+1)x_{2}}$&.&.&. &$P_{mx_{2}}$&$P_{\Delta x_{2}}$\\
$.$     & $.$        & .&.&.&.&.&.&.&.&. & . & .\\
$.$     & $.$        & .&.&.&.&.&.&.&.&. & . & .\\
$.$     & $.$        & .&.&.&.&.&.&.&.&. & . & .\\
$x_{n}$ & $P_{1x_{n}}$&$P_{2x_{n}}$&.&.&.&$P_{(k-1)x_{n}}$&$P_{(k+1)x_{n}}$&.&.&. &$P_{mx_{n}}$&$P_{\Delta x_{n}}$  \\
\hline
\end{tabular}
\end{table}

\newpage
$$ \large{\text{Potential referees}}$$

(1) Jerzy W. Grzymala-Busse, E-mail address: jerzy@ku.edu,
Affiliations: Department of Electrical Engineering and Computer
Science, University of Kansas, Lawrence, U. S. A.\vskip.10in

(2) Yiyu Yao, E-mail address: yyao@cs.uregina.ca, Affiliations:
Department of Computer Science, University of Regina, Regina,
Saskatchewan, Canada. \vskip.10in

(3) Tsau Young Lin,  E-mail address: tylin@cs.sjsu.edu,
Affiliations: Department of Computer Science, San Jose State
University, San Jose, U. S. A.\vskip.10in

(4) Weizhi Wu, E-mail address: wuwz@zjou.edu.cn, Affiliations:
School of Mathematics, Physics and Information Science, Zhejiang
Ocean University, Zhoushan, Zhejiang 316000, P. R. China.\vskip.10in

(5) Wei Yao, E-mail address: yaowei0516@163.com, Affiliations:
Department of Mathematics, Hebei University of Science and
Technology, Shijiazhuang 050018, P. R. China.\vskip.10in

(6) Ping Zhu, E-mail address: pzhubupt@gmail.com, Affiliations:
School of Science, Beijing University of Posts and
Telecommunications, Beijing 100876, P. R. China.\vskip.10in

\end{document}